\documentclass[aos,nameyear,dvips]{arximspdf}
\usepackage{dcolumn}
\usepackage{graphicx}

\doi{10.1214/10-AOS798}
\volume{38}
\issue{6}
\pubyear{2010}
\firstpage{3567}
\lastpage{3604}

\makeatletter

\newtheorem{theorem}{Theorem}
\newtheorem{corollary}{Corollary}
\newtheorem{lemma}{Lemma}
\newproclaim{remark}{Remark}
\newtheorem{proposition}{Proposition}
\newcommand{\argmin}{\mathop{\arg\min}}
\newcolumntype{d}[1]{D{.}{.}{#1}}
\newcolumntype{e}[1]{D{,}{}{#1}}

\makeatother

\begin{document}
\begin{frontmatter}

\title{Sure independence screening in generalized linear models with
NP-dimensionality\thanksref{T1}}
\runtitle{SIS in GLIM}
\thankstext{T1}{Supported in part by NSF Grants DMS-0714554 and DMS-0704337.}
\begin{aug}
\author[A]{\fnms{Jianqing} \snm{Fan}\ead[label=e1]{jqfan@princeton.edu}} \and
\author[B]{\fnms{Rui} \snm{Song}\corref{}\ead[label=e2]{song@stat.colostate.edu}}
\runauthor{J. Fan and R. Song}
\affiliation{Princeton University and Colorado State University}
\address[A]{Department of Operations Research\\
\quad and Financial Engineering\\
Princeton University\\
Princeton, New Jersey 08544\\
USA\\
\printead{e1}} %adresu isvedimo komanda gale!
\address[B]{Department of Statistics \\
Colorado State University\\
Fort Collins 80526, Colorado\\
USA\\
\printead{e2}}
\end{aug}

% HISTORY:
\received{\smonth{6} \syear{2009}}
\revised{\smonth{1} \syear{2010}}

% ABSTRACT
%
\begin{abstract}
Ultrahigh-dimensional variable selection plays an increasingly
important role in contemporary scientific discoveries and
statistical research. Among others, Fan and Lv [\textit{J. R. Stat. Soc. Ser.
B Stat. Methodol.} \textbf{70} (2008) 849--911] propose an
independent screening framework by ranking the marginal
correlations. They showed that the correlation ranking procedure
possesses a sure independence screening property within the context
of the linear model with Gaussian covariates and responses. In this
paper, we propose a more general version of the independent learning
with ranking the maximum marginal likelihood estimates or the
maximum marginal likelihood itself in generalized linear models. We
show that the proposed methods, with Fan and Lv [\textit{J. R. Stat. Soc. Ser. B Stat.
Methodol.} \textbf{70} (2008) 849--911] as a very
special case, also possess the sure screening property with
vanishing false selection rate. The conditions under which the
independence learning possesses a sure screening is surprisingly
simple. This justifies the applicability of such a simple method in
a wide spectrum. We quantify explicitly the extent to which the
dimensionality can be reduced by independence screening, which
depends on the interactions of the covariance matrix of covariates
and true parameters. Simulation studies are used to illustrate the
utility of the proposed approaches.
In addition, we establish an exponential
inequality for the quasi-maximum likelihood estimator which is
useful for high-dimensional statistical learning.
\end{abstract}

% KEYWORDS
%
\end{frontmatter}

%s1 ###
\section{Introduction}
The ultrahigh-dimensional regression problem is a significant feature in
many areas of modern scientific research using quantitative
measurements such as microarrays, genomics, proteomics, brain images
and genetic data. For example, in studying the associations
between phenotypes such as height and cholesterol\vadjust{\goodbreak} level and
genotypes, it can involve millions of SNPs; in disease
classification using microarray data, it can use thousands of
expression profiles, and dimensionality grows rapidly when
interactions are considered. Such a demand from applications brings
a lot of challenge to statistical inference, as the dimension $p$
can grow much faster than the sample size $n$ such that many models
are not even identifiable. By nonpolynomial dimensionality or
simply NP-dimensionality, we mean $\log p = O(n^{a})$ for some $a
> 0$. We will also loosely refer it to as an ultrahigh-dimensionality. The phenomenon of noise accumulation in
high-dimensional regression has also been observed by statisticians
and computer scientists. See \citet{FanLv08} and \citet
{FanFan08} for
a comprehensive review and references therein. When dimension $p$ is
ultrahigh, it is often assumed that only a small number of variables
among predictors $X_1, \ldots, X_p$ contribute to the response,
which leads to the sparsity of the parameter vector $\bolds{\beta}$.
As a
consequence, variable selection plays a prominent role in
high-dimensional statistical modeling.

Many variable selection techniques for various high-dimensional
statistical models have been proposed. Most of them are based on
the penalized pseudo-likelihood approach, such as the
bridge regression in \citet{FF93}, the LASSO in \citet
{Tibs96}, the
SCAD and other folded-concave penalty in \citet{FanLi01}, the
Dantzig selector in \citet{CT07} and their related methods
[\citet{Zou06}; \citet{ZouLi08}], to name a few.
Theoretical studies of these methods
concentrate on the persistency [\citet{Gree04}; \citet
{Geer08}], consistency
and oracle properties [\citet{FanLi01}; \citet{Zou06}].
However, in ultrahigh-dimensional statistical learning problems, these methods may not
perform well due to the simultaneous challenges of computational
expediency, statistical accuracy and algorithmic stability
[\citet{Fan08}].

\citet{FanLv08} proposed a sure independent screening (SIS)
method to
select important variables in ultrahigh-dimensional linear models.
Their proposed two-stage procedure can deal with the aforementioned
three challenges better than other methods. See also \citet{Huang08}
for a related study based on a marginal bridge regression.
\citet{FanLv08} showed that the correlation ranking of features
possesses a sure independence screening (SIS) property under certain
conditions; that is, with probability very close to 1, the
independence screening technique retains all of the important
variables in the model. However, the SIS procedure in \citet{FanLv08}
only restricts to the ordinary linear models and their technical
arguments depend heavily on the joint normality assumptions and cannot easily be extended even within the context of a linear model.
This limits significantly its use in practice which excludes
categorical variables. \citet{Huang08} also investigate the marginal
bridge regression in the ordinary linear model and their arguments
depend also heavily on the explicit expressions of the least-square
estimator and bridge regression. This calls for research on SIS
procedures in more general models and under less restrictive
assumptions.

In this paper, we consider an independence learning by ranking the
maximum marginal likelihood estimator (MMLE) or maximum marginal
likelihood itself for generalized linear models. That is, we fit $p$
marginal regressions by maximizing the marginal likelihood with
response $Y$ and the marginal covariate $X_i, i=1,\ldots,p$ (and the
intercept) each time. The magnitude of the absolute values of the
MMLE can preserve the nonsparsity information of the joint regression
models, provided that the true values of the marginal likelihood
preserve the nonsparsity of the joint regression models and that
the MMLE estimates the true values of the marginal likelihood
uniformly well. The former holds under a surprisingly simple
condition, whereas the latter requires a development of uniform
convergence over NP-dimensional marginal likelihoods. \citet{HTX09}
used a different marginal utility,
derived from an empirical likelihood point of view. \citet{HM09}
proposed a generalized correlation ranking, which
allows nonlinear regression. Both papers proposed an interesting
bootstrap method to assess the authority of the selected features.

As the MMLE or maximum likelihood ranking is equivalent to the
marginal correlation ranking in the ordinary linear models, our work
can thus be considered as an important extension of SIS in
\citet{FanLv08}, where the joint normality of the response and
covariates is imposed. Moreover, our results improve over those in
\citet{FanLv08} in at least three aspects. First, we establish a
new framework for having SIS properties, which does not build on the
normality assumption even in the linear model setting. Second,
while it is not obvious (and could be hard) to generalize the proof
of \citet{FanLv08} to more complicated models, in the current
framework, the SIS procedure can be applied to the generalized
linear models and possibly other models. Third, our results can
easily be applied to the generalized correlation ranking
[\citet{HM09}] and other rankings based on a group of marginal
variables.

Fitting marginal models to a joint regression is a type of model
misspecification [\citet{Whit82}], since we drop out most covariates
from the model fitting. In this paper, we establish a
nonasymptotic tail probability bound for the MMLE under model
misspecifications, which is beyond the traditional asymptotic
framework of model misspecification and of interest in its own
right. As a practical screening method, independent screening can
miss variables that are marginally weakly correlated with the
response variables, but jointly highly important to the response
variables, and also rank some jointly unimportant variables too
high by using marginal methods. \citet{FanLv08} and \citet{Fan08}
develop iteratively conditional screening and selection
methods to make the procedures robust and practical. The former
focuses on ordinary linear models and the latter improves the idea
in the former and expands significantly the scope of
applicability, including generalized linear models.

The SIS property can be achieved as long as the surrogate, in this
case, the marginal utility, can preserve the nonsparsity of the
true parameter values. With a similar idea, \citet{Fan08}
proposed a SIS procedure for generalized linear models, by sorting
the maximum likelihood functions,
which is a type of ``marginal likelihood ratio'' ranking, whereas the
MMLE can be viewed as a Wald type of statistic. The two
methods are equivalent in terms of sure screening properties in
our proposed framework. This will be demonstrated in our paper.
The key technical challenge in the maximum marginal likelihood
ranking is that the signal can even be weaker than the noise.
We overcome this technical difficulty by using the invariance
property of ranking under monotonic transforms.

The rest of the paper is organized as follows. In Section~\ref{sec2}, we
briefly introduce the setups of the generalized linear models. The
SIS procedure is presented in Section~\ref{sec3}. In Section~\ref{sec4}, we provide
an exponential bound for quasi maximum likelihood estimator. The
SIS properties of the MMLE learning are presented in Section~\ref{sec5}. In
Section~\ref{sec6}, we formulate the marginal likelihood screening and show
the SIS property. Some simulation results are presented in Section~\ref{sec7}.
A~summary of our findings and discussions is in
Section~\ref{sec8}. The detailed proofs are relegated to Section~\ref{sec9}.

%s2 ###
\section{Generalized linear models}\label{sec2}
%In our definition of generalized linear models, we will follow the
%notation of \citet{McCullagh89}.
Assume that the random scalar $Y$
is from an exponential family with the probability density
function taking the canonical form
%e1 ###
%
\begin{eqnarray}\label{e1}
f_Y(y; \theta) = \exp\{y\theta- b(\theta) + c(y) \}
\end{eqnarray}
for some known functions $b(\cdot)$, $c(\cdot)$ and unknown
function $\theta$. Here we do not consider the dispersion
parameter as we only model the mean regression. We can easily
introduce a dispersion parameter in (\ref{e1}) and the results
continue to hold. The function $\theta$ is usually called the
canonical or natural parameter. The mean response is $b'(\theta)$,
the first derivative of $b(\theta)$ with respect to $\theta$.
We consider the problem of estimating a $(p+1)$-vector
of parameter $\bolds{\beta}= (\beta_0,\beta_1,\ldots, \beta_p)$
from the
following generalized linear model:
%e2 ###
%
\begin{eqnarray}\label{e2}
E(Y|\mathbf{X}= \mathbf x) = b'(\theta(\mathbf{x})) =
g^{-1}\Biggl(\sum_{j=0}^p \beta_j x_j\Biggr),
\end{eqnarray}
where $\mathbf{x} = \{x_0,x_1,\ldots, x_p\}^T$ is a
($p+1$)-dimensional covariate and $x_0=1$ represents the intercept.
If $g$ is the canonical link, that is, $g = (b')^{-1}$, then
$\theta(x) = \sum_{j=0}^p \beta_j x_j$. We focus on the canonical
link function in this paper for simplicity of presentation.

Assume that the observed data $\{(\mathbf{X}_i, Y_i), i=1,\ldots,n\}$ are
i.i.d. copies of $(\mathbf{X},Y)$, where the covariate $\mathbf{X}= (X_0,
X_1,\ldots,X_p)$ is a $(p+1)$-dimensional random vector and $X_0 =
1$. We allow $p$ to grow with $n$ and denote it as $p_n$ whenever
needed.

We note that the ordinary linear model $Y = \mathbf{X}^T \bolds{\beta}+
\varepsilon$ with $\varepsilon\sim N(0, 1)$ is a special case of
model (\ref{e2}), by taking $g(\mu) = \mu$ and $b(\theta) =
\theta^2/2$. When the design matrix $\mathbf{X}$ is standardized, the
ranking by the magnitude of the marginal correlation is in fact the
same as the ranking by the magnitude of the
maximum marginal likelihood estimator (MMLE). Next we
propose an independence screening method to GLIM based on the MMLE.
We also assume that the covariates are standardized to have mean
zero and standard deviation one
\[
EX_j = 0 \quad\mbox{and}\quad EX_j^2 = 1,\qquad j = 1,\ldots,p_n.
\]

%s3 ###
\section{Independence screening with MMLE}\label{sec3}
Let $\mathcal{M}_{\star} = \{1 \le j \le p_n\dvtx  \beta_j^{\star} \neq
0 \}$ be the true sparse model with nonsparsity size $s_n =
|\mathcal{M}_{\star}|$, where $\bolds{\beta}^{\star} = (\beta
_0^{\star},
\beta_1^{\star}, \ldots, \beta_{p_n}^{\star})$ denotes the true
value. In this paper, we refer to marginal models as fitting models
with componentwise covariates. The maximum marginal likelihood
estimator (MMLE) $\hat{\bolds{\beta}}{}^M_{j}$, for $j=1,\ldots,
p_n$, is defined as the minimizer of the componentwise regression
\begin{eqnarray*}
\hat{\bolds{\beta}}{}^M_{j} = (\hat{\beta}_{j,0}^M, \hat{\beta
}_{j}^M )
= \mathop{\arg\min}_{\beta_0, \beta_j } \mathbb{P}_n l(
\beta_0+\beta_jX_{j},Y),
\end{eqnarray*}
where $l(Y;\theta) = - [\theta Y - b(\theta) - \log
c(Y) ]$ and $\mathbb{P}_n f(X, Y) = n^{-1}\sum_{i=1}^n f(X_i,
Y_i)$ is the empirical measure. This can be rapidly computed and
its implementation is robust, avoiding numerical instability in
NP-dimensional problems. We correspondingly define the population
version of the minimizer of the componentwise regression,
\begin{eqnarray*}
{\bolds{\beta}}_{j}^M = ({\beta}_{j,0}^M, {\beta}_{j}^M ) =
\mathop{\arg\min}_{\beta_0, \beta_j } {E} l( \beta_0+\beta_jX_{j},Y)
\qquad\mbox{for } j=1,\ldots, p_n,
\end{eqnarray*}
where $E$ denotes the expectation under the true model.

We select a set of variables
%e3 ###
%
\begin{eqnarray}\label{e3}
\widehat{\mathcal{M}}_{\gamma_n} = \{1 \le j \le p_n\dvtx  |\hat
\beta_{j}^M| \ge\gamma_n \},
\end{eqnarray}
where $\gamma_n$ is a predefined threshold value. Such an
independence learning ranks the importance of features according to
their magnitude of marginal regression coefficients. With an
independence learning, we dramatically decrease the dimension of the
parameter space from $p_n$ (possibly hundreds of thousands) to a
much smaller number by choosing a large $\gamma_n$, and hence the
computation is much more feasible. Although the interpretations and
implications of the marginal models are biased from the joint model,
the nonsparse information about the joint model can be passed along
to the marginal model under a mild condition. Hence it is suitable
for the purpose of variable screening. Next we will show under
certain conditions that the sure screening property holds, that is, the
set $\mathcal{M}_{\star}$ belongs to
$\widehat{\mathcal{M}}_{\gamma_n}$ with probability one
asymptotically, for an appropriate choice of $\gamma_n$. To
accomplish this, we need the following technical device.

%s4 ###
\section{An exponential bound for QMLE}\label{sec4}
In this section, we obtain an exponential bound for the quasi-MLE
(QMLE), which will be used in the next section. Since this result
holds under very general conditions and is of self-interest, in
the following we make a more general description of the model and
its conditions.

Consider data $\{\mathbf{X}_i, Y_i\}$, $i=1,\ldots,n,$ are $n$ i.i.d. samples
of $(\mathbf{X},Y) \in\mathcal{X} \times\mathcal{Y}$ for some space
$\mathcal{X}$ and $\mathcal{Y}$. A regression model for $\mathbf{X}$ and
$Y$ is assumed with quasi-likelihood function $-l(\mathbf{X}^T \bolds
{\beta},
Y)$. Here $Y$ and $\mathbf{X}= (X_1, \ldots, X_q)^T$ represent the
response and the $q$-dimensional covariate vector, which may
include both discrete and continuous components and the
dimensionality can also depend on $n$. Let
\[
\bolds{\beta}_0 = \mathop{\arg\min}_{\bolds{\beta}} E l(\mathbf
{X}^T\bolds{\beta}, Y)
\]
be the population parameter. Assume that $\bolds{\beta}_0$ is an
interior point of a
sufficiently large, compact and convex set $\mathbf{B}\in\mathbf R^q$.
The following conditions on the model are needed:
\begin{enumerate}[(A)]
\item[(A)] The Fisher information,
\begin{eqnarray*}
I(\bolds{\beta}) = E \biggl\{ \biggl[ \frac{\partial}{\partial\bolds{\beta}} l(
\mathbf{X}^T \bolds{\beta}, Y) \biggr] \biggl[\frac{\partial}{\partial\bolds
{\beta}} l(
\mathbf{X}^T \bolds{\beta}, Y) \biggr]^T \biggr\},
\end{eqnarray*}
is finite and positive definite at $\bolds{\beta}= \bolds{\beta}_0$.
Moreover, $\|I(\bolds{\beta}) \|_{\mathbf{B}} =$ \\
$\sup_{\bolds{\beta}\in\mathbf{B}, \|\mathbf{x}\| = 1}\|I(\bolds
{\beta})^{1/2}
\mathbf{x}\| $
exists, where $\|\cdot\|$ is the Euclidean norm.
\item[(B)] The function $l(\mathbf x^{T} \bolds{\beta},y)$
satisfies the Lipschitz property with positive constant $k_n$
\[
|l(\mathbf x^T \bolds{\beta}, y) - l(\mathbf x^T \bolds{\beta}',y)
|I_n(\mathbf
{x}, y)
\le k_n|\mathbf x^T \bolds{\beta}-\mathbf x^T \bolds{\beta
}'|I_n(\mathbf{x},
y)
\]
for $\bolds{\beta}, \bolds{\beta}' \in\mathbf{B}$, where $
I_n(\mathbf{x}, y)
=I((\mathbf{x}, y)\in\Omega_n)$ with
\[
\Omega_n =\{(\mathbf{x}, y)\dvtx  \|\mathbf{x}\|_\infty\leq K_n,
|y|\leq K_n^{\star}\}
\]
for some sufficiently large positive
constants $K_n$ and $K_n^{\star}$, and $\|\cdot\|_\infty$ being
the supremum norm. In addition, there exists a sufficiently large
constant $C$ such
that with $b_n = C k_nV_n^{-1} (q/n)^{1/2}$ and $V_n$ given in
condition~C
\[
\sup_{\bolds{\beta}\in\mathbf{B},\|\bolds{\beta}- \bolds{\beta
}_0 \|\le b_n }
| E [l(\mathbf{X}^T \bolds{\beta}, Y) - l(\mathbf{X}^T \bolds
{\beta}_0, Y)](1-I_n(\mathbf{X}, Y)) |
\le o(q/n),
\]
where $V_n$ is the constant given in condition~C.
\item[(C)] The function $l(\mathbf{X}^T\bolds{\beta}, Y)$ is convex
in $\bolds{\beta}$,
satisfying
\[
E\bigl(l(\mathbf\mathbf{X}^T \bolds{\beta}, Y) -
l(\mathbf\mathbf{X}^T \bolds{\beta}_0, Y)\bigr) \ge V_n \|\bolds{\beta}-
\bolds{\beta}_0 \|^2
\]
for all $\|\bolds{\beta}-\bolds{\beta}_0\| \leq b_n$ and some
positive constants $V_n$.
\end{enumerate}

Condition A is analogous to assumption A6(b) of \citet{Whit82} and
assumption $R_s$ in \citet{Fahrmeir86}. It ensures the
identifiability and the existence of the QMLE and is satisfied
for many examples of generalized linear models. Conditions A and C
are overlapped but not the same.
%The constants $k$ and $V$ can depend on $n$.

We now establish an exponential bound for the tail probability of
the QMLE
\[
\hat{\bolds{\beta}} = \mathop{\arg\min}_{\bolds{\beta}} \mathbb{P}_n
l( \mathbf{X}^T\bolds{\beta},Y).
\]
The idea of the proof is to connect $\sqrt{n}\|\hat{\bolds{\beta}} -
{\bolds{\beta}}_0\|$ to the tail of certain empirical processes and
utilize the convexity and Lipschitz continuities.

\begin{theorem}\label{the-1}
Under conditions \textup{A--C}, it holds that for any $t > 0$,
\begin{eqnarray*}
P \bigl(\sqrt{n}\| \hat{\bolds{\beta}} - \bolds{\beta}_0\|
\ge16k_n (1+t)/V_n \bigr)
\le
\exp(-2 t^2/K_n^2 ) + n P ( \Omega_n^c).
\end{eqnarray*}
\end{theorem}

%s5 ###
\section{Sure screening properties with MMLE}\label{sec5}
In this section, we introduce a new framework for establishing the
sure screening property with MMLE in the canonical exponential
family (\ref{e1}). We divide into three sections to present our
findings.

%s5.1 ###
\subsection{Population aspect}
As fitting marginal regressions to a joint regression is a type of
model misspecification, an important question would be: at what
level the model information is preserved. Specifically for
screening purposes, we are interested in the preservation of the
nonsparsity from the joint regression to the marginal regression.
This can be summarized into the following two questions. First,
for the sure screening purpose, if a variable $X_j$ is jointly
important $(\beta_{j}^{\star} \neq0)$, will (and under what
conditions) it still be marginally important $(\beta_{j}^M \neq
0)$? Second, for the model selection consistency purpose, if a
variable $X_j$ is jointly unimportant $(\beta_{j}^{\star} = 0)$,
will it still be marginally unimportant $(\beta_{j}^M = 0)$? We
aim to answer these two questions in this section.

The following theorem reveals that the marginal regression
parameter is in fact a measurement of the correlation between the
marginal covariate and the mean response function.
%t2
\begin{theorem}\label{the-2}
For $j=1,\ldots, p_n$, the marginal regression parameters
$\beta_j^M =0$ if and only if $\operatorname{cov}(b'(\mathbf{X}^T
\bolds{\beta}^{\star}), X_j)
= 0$.
\end{theorem}

By using the fact that that $\mathbf{X}^T \bolds{\beta}^{\star} =
\beta_0^{\star} + \sum_{j \in\mathcal{M}_{\star}}X_j
\beta_j^{\star}$, we can easily show the following corollary.\vadjust{\goodbreak}
%cor1
\begin{corollary}\label{cor-1}
If the partial orthogonality condition holds, that is, $\{X_j,\break j
\notin\mathcal{M}_{\star}\}$ is independent of $\{ X_i, i \in
\mathcal{M}_{\star}\}$, then $\beta_j^M =0$, for $j \notin
\mathcal{M}_{\star}.$
\end{corollary}

This partial orthogonality condition is essentially the assumption
made in \citet{Huang08} who showed the model selection
consistency in the special case with the ordinary linear model and
bridge regression. Note that $\operatorname{cov}(b'(\mathbf{X}^T
\bolds{\beta}^{\star}), X_j)
= \operatorname{cov}(Y, X_j)$. A necessary condition for sure
screening is that
the important variables $X_j$ with $\beta_j^{\star} \neq0$ are
correlated with the response, which usually holds. When they are
correlated with the response, by Theorem~\ref{the-2}, $\beta_j^M
\neq0$, for $j \in\mathcal{M}_{\star}.$ In other words, the
marginal model pertains to the information about the important
variables in the joint model. This is the theoretical basis for
the sure independence screening. On the other hand, if the partial
orthogonality condition in Corollary 1 holds, then $\beta_j^M=0$
for $j \notin\mathcal{M}_{\star}$. In this case, there exists a
threshold $\gamma_n$ such that the marginally selected model is
model selection consistent
\[
\min_{j \in\mathcal{M}_{\star}} |\beta_{j}^M| \geq\gamma_n,\qquad
\max_{j \notin\mathcal{M}_{\star}} |\beta_{j}^M| = 0.
\]

To have a sure screening property based on the sample version
(\ref{e3}), we need
\[
\min_{j \in\mathcal{M}_{\star}}|\beta_j^M|\geq O(n^{-\kappa})
\]
for some $\kappa< 1/2$ so that the marginal signals are stronger
than the stochastic noise. The following theorem shows that this
is possible.
%t3
\begin{theorem}\label{the-3}
If $|\operatorname{cov}(b'(\mathbf{X}^T \bolds{\beta}^{\star}),
X_j)| \ge c_1
n^{-\kappa}$ for $j \in\mathcal{M}_{\star}$ and a positive constant
$c_1 > 0$, then
there exists a positive constant $c_2$ such that
\[
\min_{j \in\mathcal{M}_{\star}}|\beta_j^M| \ge c_2 n^{-\kappa},
\]
provided that $b''(\cdot)$ is bounded or
\[
E G(a|X_j|) |X_j| I(|X_j| \geq n^{\eta}) \leq d n^{-\kappa}\qquad\mbox{for some }
 0<\eta<\kappa,
\]
and some sufficiently small positive constants $a$ and $d$, where
$G(|x|) = \break \sup_{|u| \leq|x|} | b'(u)|$.
\end{theorem}

Note that for the normal and Bernoulli distribution, $b''(\cdot)$
is bounded, whereas for the Poisson distribution, $G(|x|) =
\exp(|x|)$ and Theorem~\ref{the-3} requires the tails of $X_j$ to be light.
Under some additional conditions, we will show in the proof of
Theorem~\ref{the-5} that
\[
\sum_{j=1}^p |\beta_j^M|^2 = O(\| \bolds{\Sigma}\bolds{\beta
}^\star\|^2) =
O(\lambda_{\max} (\bolds{\Sigma})),
\]
where $\bolds{\Sigma}= \operatorname{var}(\mathbf{X})$,
and $\lambda_{\max} (\bolds{\Sigma})$ is its maximum eigenvalue. The
first equality requires some efforts to prove, whereas the second
equality follows easily from the assumption
\[
\operatorname{var}(\mathbf{X}^T\bolds{\beta}^\star) = \bolds
{\beta}^\star{}^T \bolds{\Sigma}
\bolds{\beta}^\star=O(1).
\]
The implication of this result is that there cannot be too many
variables that have marginal coefficient $|\beta_j^M|$ that
exceeds certain thresholding level. That achieves the sparsity in
final selected model.

When the covariates are jointly normally distributed, the
condition of Theorem~\ref{the-3} can be further simplified.
%p1
\begin{proposition} \label{prop1}
Suppose that $X$ and $Z$ are jointly normal with mean zero and
standard deviation 1. For a strictly monotonic function
$f(\cdot)$,\break $\operatorname{cov}(X, Z) = 0$ if and only if $\operatorname{cov}(X, f(Z)) = 0$,
provided the latter covariance exists. In addition,
\[
|\operatorname{cov}(X, f(Z))| \geq|\rho| \inf_{|x| \leq c|\rho|} |g'(x)|
EX^2 I(|X| \leq c)
\]
for any $c > 0$, where $\rho= E X Z$, $g(x) = E f(x +
\varepsilon)$ with $\varepsilon\sim N(0, 1-\rho^2)$.
\end{proposition}

The above proposition shows that the covariance of $X$ and $f(Z)$
can be bounded from below by the covariance between $X$ and $Z$,
namely
\[
|\operatorname{cov}(X, f(Z))| \geq d |\rho|,\qquad
d = \inf_{|x| \leq c}|g'(x)| EX^2 I(|X| \leq c),
\]
in which $d > 0$ for a sufficiently small $c$. The first part of the
proposition actually holds when the conditional density $f(z|x)$ of
$Z$ given $X$ is a monotonic likelihood family [\citet{BD01}]
when $x$
is regarded as a parameter. By taking $Z = \mathbf{X}^T \bolds{\beta
}^\star$, a
direct application of Theorem~\ref{the-2} is that $\beta_{j}^M = 0$
if and only if
\[
\operatorname{cov}(\mathbf{X}^T \bolds{\beta}^\star, X_j) = 0,
\]
provided that $\mathbf{X}$ is jointly normal, since $b'(\cdot)$ is an
increasing function. Furthermore, if
%e4 ###
%
\begin{equation}\label{e4}
|\operatorname{cov}(\mathbf{X}^T \bolds{\beta}^\star, X_j)| \geq
c_0 n^{-\kappa},\qquad \kappa< 1/2,
\end{equation}
for some positive constant $c_0$, a minimum condition required
even for the least-squares model [\citet{FanLv08}], then by the
second part of Proposition~\ref{prop1}, we have
\[
|\operatorname{cov}(b'(\mathbf{X}^T \bolds{\beta}^{\star}), X_j)|
\geq c_1 n^{-\kappa}
\]
for some constant $c_1$. Therefore, by Theorem~\ref{the-2}, there
exists a positive constant $c_2$ such that
\[
|\beta_j^M| \geq c_2 n^{-\kappa}.
\]
In other words, (\ref{e4}) suffices to have marginal signals that
are above the maximum noise level.

%s5.2 ###
\subsection{Uniform convergence and sure screening}\label{sec5.2}
To establish the SIS property of MMLE, a key point is to establish
the uniform convergence of the MMLEs. That is, to control the
maximum noise level relative to the signal. Next we establish the uniform
convergence rate for the MMLEs and sure screening property of the
method in (\ref{e3}). The former will be useful in controlling the
size of the selected set.

Let $\bolds{\beta}_j=(\beta_{j,0}, \beta_{j})^T$ denote the
two-dimensional parameter and $\mathbf{X}_j = (1, X_j)^T$. Due to the
concavity of the log-likelihood in GLIM with the canonical link,
$E l(\mathbf{X}_j^T \bolds{\beta}_j,Y)$ has a unique minimum over
$\bolds{\beta}_j \in
\mathcal{B}$ at an interior point $\bolds{\beta}_{j}^{M} = (\beta_{j,0}^M,
\beta_{j}^M)^T$, where $\mathcal{B} = \{|\beta_{j,0}^M| \leq B,
|\beta_j^M| \leq B\}$ is a square with the width $B$
over which the marginal likelihood is maximized. The
following is an updated version of conditions~A--C for each
marginal regression and two additional conditions for the
covariates and the population parameters:
\begin{enumerate}[A$'$.]
\item[A$'$.] The marginal Fisher information: $I_j(\bolds{\beta}_j) =
E \{ b''(\mathbf{X}_j^T\bolds{\beta}_j) \mathbf{X}_j \mathbf
{X}_j^T \}$ is finite and
positive definite at $\bolds{\beta}_j = \bolds{\beta}_j^M$, for
$j=1,\ldots,p_n$. Moreover, $\|I_j(\bolds{\beta}_j) \|_{\mathcal
{B}}$ is bounded
from above.
\item[B$'$.] The second derivative of $b(\theta)$ is
continuous and positive. There exists an $\varepsilon_1>0$ such
that for all $j=1,\ldots,p_n$,
\[
\sup_{\bolds{\beta}\in\mathcal{B},~\|\bolds{\beta}- \bolds
{\beta}_j^M \|\le
\varepsilon_1}
|Eb(\mathbf{X}_j^T \bolds{\beta}) I(|X_j| > K_n)| \le o(n^{-1}).
\]

\item[C$'$.] For all $\bolds{\beta}_j \in\mathcal{B}$, we have
$E(l(\mathbf{X}_j^T \bolds{\beta}_j, Y) - l(\mathbf{X}_j^T \bolds
{\beta}_{j}^M, Y))
\ge V \|\bolds{\beta}_j - \bolds{\beta}_{j}^M \|^2 ,$ for some positive
$V$, bounded from below uniformly over $j=1,\ldots,p_n$.
\item[D.] There exists some positive constants
$m_0$, $m_1$, $s_0$, $s_1$ and $\alpha$, such that for sufficiently
large $t$,
\[
P(|X_j|> t) \le(m_1 - s_1)\exp\{-m_{0}t^{\alpha}\}
\qquad\mbox{for $j=1,\ldots,p_n$},
\]
and that
\[
E \exp\bigl(b(\mathbf{X}^T \bolds{\beta}^{\star}+s_0) - b(\mathbf{X}^T
\bolds{\beta}^{\star})\bigr)+
E \exp\bigl(b(\mathbf{X}^T \bolds{\beta}^{\star}-s_0) - b(\mathbf{X}^T
\bolds{\beta}^{\star})\bigr)
\le s_1.
\]

\item[E.] The conditions in Theorem~\ref{the-3} hold.
%and $\min_{j \in\mathcal{M}_*} |\beta_j^\star| \geq c_3 n^{-\kappa} $
%for a
%sufficiently large constant $c_3$.
\end{enumerate}

Conditions A$'$--C$'$ are satisfied in a lot of examples of generalized
linear models, such as linear regression, logistic regression and
Poisson regression. Note that the second part of condition~D ensures
the tail of the response variable $Y$
to be exponentially light, as shown in the following lemma:
%l1
\begin{lemma}\label{lem-1}
If condition~D holds, for any $t>0$,
\[
P(|Y| \ge m_0 t^{\alpha}/s_0) \le s_1\exp(-m_0 t^{\alpha}).
\]
\end{lemma}

Let $k_n = b'(K_n B + B) + m_0K_n^{\alpha}/s_0$. Then condition~B
holds for exponential family (\ref{e1}) with $K_n^{\star} = m_0
K_n^{\alpha}/s_0$. The Lipschitz constant $k_n$ is bounded for the
logistic regression, since $Y$ and $b'(\cdot)$ are bounded. The
following theorem gives a uniform\vadjust{\goodbreak} convergence result of MMLEs and a
sure screening property. Interestingly, the sure screening property
does not directly depend on the property of the covariance matrix of
the covariates such as the growth of its operator norm. This is an
advantage over using the full likelihood.
%t4
\begin{theorem}\label{the-4}
Suppose that conditions~\textup{A$'$, B$'$, C$'$} and \textup{D} hold.
\begin{enumerate}[(ii)]
\item[(i)] If $n^{1-2\kappa}/(k_n^2K_n^2) \to\infty$, then for
any $c_3 >0$, there exists a positive constant $c_4$ such that
\begin{eqnarray*}
& & P \Bigl(\max_{1 \le j \le p_n}|\hat\beta_{j}^M - \beta_{j}^M|
\ge
c_3 n^{-\kappa} \Bigr) \\
&&\qquad \leq p_n \bigl\{ \exp\bigl( - c_4 n^{1-2\kappa} /(k_nK_n)^2\bigr) +
nm_1\exp(-m_0 K_n^\alpha)\bigr\}.
\end{eqnarray*}

\item[(ii)] If, in addition, condition~E holds, then by taking
$\gamma_n = c_5 n^{-\kappa}$ with $c_5 \leq c_2/2$, we have
\begin{eqnarray*}
P (\mathcal{M}_{\star} \subset\widehat
{\mathcal{M}}_{\gamma_n} ) \ge1 - s_n \bigl\{ \exp\bigl( - c_4
n^{1-2\kappa} /(k_nK_n)^2\bigr) + nm_1\exp(-m_0 K_n^\alpha)\bigr\},
\end{eqnarray*}
\end{enumerate}
where $s_n = |\mathcal{M}_\star|$, the size of nonsparse
elements.
\end{theorem}
%r1
\begin{remark}
If we assume that $\min_{j \in\mathcal{M}_*} |\operatorname
{cov}(b'(\mathbf{X}^T
\bolds{\beta}^\star), X_j)| \geq c_1 n^{-\kappa+\delta}$ for any
$\delta
>0$, then one can take $\gamma_n = c n^{-\kappa+\delta/2}$ for any
$c > 0$ in Theorem~\ref{the-4}. This is essentially the
thresholding used in Fan and Lv (\citeyear{FanLv08}).
\end{remark}

Note that when $b'(\cdot)$ is bounded as the Bernoulli model,
$k_n$ is a finite constant. In this case, by balancing the two
terms in the upper bound of Theorem~\ref{the-4}(i), the optimal
order of $K_n$ is given by
\[
K_n = n^{(1-2\kappa)/(\alpha+2)}
\]
and
\[
P \Bigl(\max_{1 \le j \le p_n}|\hat\beta_{j}^M - \beta_{j}^M| \ge
c_3 n^{-\kappa} \Bigr) = O \bigl\{ p_n\exp\bigl( -c_4
n^{(1-2\kappa)\alpha/(\alpha+2)}\bigr) \bigr\},
\]
for a positive constant $c_4$. When the covariates $X_j$ are
bounded, then $k_n$ and $K_n$ can be taken as finite constants.
In this case,
\[
P \Bigl(\max_{1 \le j \le p_n}|\hat\beta_{j}^M - \beta_{j}^M| \geq
c_3 n^{-\kappa} \Bigr)
\leq O \{ p_n \exp( - c_4 n^{1-2\kappa}) \}.
\]
In both aforementioned cases, the tail probability in
Theorem~\ref{the-4} is exponentially small. In other words, we
can handle the NP-dimensionality,
\[
\log p_n = o \bigl(n^{(1-2\kappa)\alpha/(\alpha+2)} \bigr),
\]
with $\alpha=\infty$ for the case of bounded covariates.

For the ordinary linear model, $k_n = B(K_n+1)+K_n^{\alpha}/(2s_0)$
and by taking
the optimal order of $K_n = n^{(1-2\kappa)/A}$ with $A =
\max(\alpha+4, 3\alpha+2)$, we have
\[
P \Bigl(\max_{1 \le j \le p_n}|\hat\beta_{j}^M - \beta_{j}^M| > c_3
n^{-\kappa} \Bigr) = O
\bigl\{ p_n\exp\bigl( -c_4 n^{(1-2\kappa)\alpha/A}\bigr)
\bigr\}.
\]
When the covariates are normal, $\alpha= 2$ and our result is
weaker than that given in Fan and Lv (\citeyear{FanLv08}) who permits $\log p_n
= o (n^{1-2\kappa})$ whereas Theorem~\ref{the-4} can only handle
$\log p_n = o (n^{(1-2\kappa)/4})$. However, we allow nonnormal
covariate and other error distributions.

The above discussion applies to the sure screening property given
in Theorem~\ref{the-4}(ii). It is only the size of nonsparse
elements $s_n$ that matters for the purpose of sure screening,
not the dimensionality $p_n$.

%s5.3 ###
\subsection{Controlling false selection rates}
After applying the variable screening procedure, the question
arrives naturally how large the set $\widehat
{\mathcal{M}}_{\gamma_n}$ is. In other words, has the number of
variables been actually reduced by the independence learning? In
this section, we aim to answer this question.

A simple answer to this question is the ideal case in which
\[
\operatorname{cov}(b'(\mathbf{X}^T \bolds{\beta}^{\star}), X_j)
=o(n^{-\kappa})\qquad\mbox{for } j \notin\mathcal{M}_{\star}.
\]
In this case, under some mild conditions, we can show (see the proof of
Theorem~\ref{the-3}) that
\[
\max_{j \notin\mathcal{M}_{\star}} |\beta_j^M| =
o(n^{-\kappa}).
\]
This, together with Theorem~\ref{the-4}(i) shows that
\[
\max_{j \notin\mathcal{M}_{\star}} |\hat{\beta}_j^M| \leq c_3
n^{-\kappa}\qquad \mbox{for any $c_3 > 0$,}
\]
with probability tending to one if the probability in
Theorem~\ref{the-4}(i) tends to zero. Hence, by the choice of
$\gamma_n$ as in Theorem~\ref{the-4}(ii), we can achieve model
selection consistency
\[
P( \widehat
{\mathcal{M}}_{\gamma_n} = \mathcal{M}_{\star} ) = 1 - o(1).
\]
This kind of condition was indeed implied by the condition in
\citet{Huang08} in the special case with ordinary linear model
using the bridge regression who draw a similar conclusion.

We now deal with the more general case. The idea is to bound the
size of the selected set (\ref{e3}) by using the fact $\operatorname
{var}(Y)$ is
bounded. This usually implies $\operatorname{var}(\mathbf{X}^T \bolds
{\beta}^\star) =
\bolds{\beta}^\star{}^T \bolds{\Sigma}\bolds{\beta}^\star=
O(1)$. We need the following
additional conditions:
\begin{enumerate}[G.]
\item[F.] The variance $\mbox{var}(\mathbf{X}^T \bolds{\beta
}^{\star})$ is
bounded from
above and below.

\item[G.] Either $b''(\cdot)$ is bounded or $\mathbf{X}_M
= (X_1,\ldots,X_{p_n})^T$ follows an
elliptically contoured distribution, that is,
\[
\mathbf{X}_M = \bolds{\Sigma}_1^{1/2} R \mathbf{U},
\]
and $|E b'(\mathbf{X}^T \bolds{\beta}^\star) (\mathbf{X}^T \bolds
{\beta}^\star-
\beta_0^{\star})|$ is bounded, where $\mathbf{U}$ is uniformly
distributed on the unit sphere in $p$-dimensional Euclidean space,
independent of the nonnegative random variable $R$, and $\bolds{\Sigma
}_1 =
\operatorname{var}(\mathbf{X}_M).$
\end{enumerate}

Note that $\bolds{\Sigma}= \operatorname{diag}(0, \bolds{\Sigma}_1)$ in
condition~G$'$,
since the covariance matrices differ only in the intercept term. Hence,
$\lambda_{\max}(\bolds{\Sigma}) = \lambda_{\max}(\bolds{\Sigma}_1)$.
The following\vspace*{1pt} result is about the size of $\widehat
{\mathcal M}_{\gamma_n}$.
%t5
\begin{theorem}\label{the-5}
Under conditions~\textup{A$'$, B$'$, C$'$, D, F} and \textup{G}, we have for any $\gamma_n =
c_5 n^{-2 \kappa}$, there exists a $c_4$ such that
\begin{eqnarray*}
&& P [ | \widehat{\mathcal{M}}_{\gamma_n}| \leq
O\{n^{2\kappa}\lambda_{\max} (\bolds{\Sigma})\} ] \\
&&\qquad \geq 1 - p_n \bigl\{ \exp\bigl( - c_4 n^{1-2\kappa} /(k_nK_n)^2\bigr) +
nm_1\exp(-m_0 K_n^\alpha)\bigr\}.
\end{eqnarray*}
\end{theorem}

The right-hand side probability has been explained in Section~\ref{sec5.2}.
From the proof of Theorem~\ref{the-5}, we actually show that the
number of selected variables is of order $\|\bolds{\Sigma}
\bolds{\beta}^\star\|^2/\gamma_n^2$, which is further bounded by
$O\{n^{2\kappa}\lambda_{\max} (\bolds{\Sigma})\}$ using
$\operatorname{var}(\mathbf{X}^T\bolds{\beta}^\star) = O(1)$.
Interestingly, while the sure
screening property does not depend on the behavior of $\bolds{\Sigma
}$, the
number of selected variables is affected by how correlated the
covariates are. When $n^{2\kappa}\lambda_{\max} (\bolds{\Sigma}) /
p \to
0$, the number of selected variables are indeed negligible comparing
to the original size. In this case, the percent of falsely
discovered variables is of course negligible. In particular, when
$\lambda_{\max} (\bolds{\Sigma}) = O(n^{\tau})$, the size of selected
variable is of order $O(n^{2\kappa+\tau})$. This is of the same
order as in Fan and Lv (\citeyear{FanLv08}) for the multiple regression model with
the Gaussian data who needs additional condition that $2 \kappa+
\tau< 1$. Our result is an extension of Fan and Lv (\citeyear{FanLv08}) even in
this very specific case without the condition $2 \kappa+ \tau< 1$.
In addition, our result is more intuitive: the number of selected
variables is related to $\lambda_{\max} (\bolds{\Sigma})$, or, more
precisely, $\|\bolds{\Sigma}\bolds{\beta}^{\star} \|^2$ and the
thresholding
parameter $\gamma_n$.

%s6 ###
\section{A likelihood ratio screening}\label{sec6}
In a similar variable screening problem with generalized linear
models, \citet{Fan08} suggest to screen the variables by
sorting the marginal likelihood. This method can be viewed as a
marginal likelihood ratio screening, as it builds on the increments of the
log-likelihood. In this section we show that the likelihood ratio
screening is equivalent to the MMLE screening
in the sense that they both possess the sure screening property
and that the number of selected variables of the two methods are
of the same order of magnitude.

We first formulate the marginal likelihood screening procedure.
Let
\[
L_{j,n} = \mathbb{P} _n \{l(\hat{\beta}_0^M,Y) - l(\mathbf{X}_j^T
\hat{\bolds{\beta}}{}^M_j,
Y) \},\qquad j=1,\ldots,p_n,
\]
and $\mathbf{L}_n = (L_{1,n},\ldots, L_{p_n,n})^T$, where $\hat
{\beta}_0^M =
\argmin_{\beta_0} \mathbb{P} _n l(\beta_0, Y)$. Correspondingly, let
\[
L_{j}^{\star} = E \{l(\beta_0^M,Y) - l(\mathbf{X}_j^T\bolds{\beta}_j^M,
Y) \},\qquad j=1,\ldots,p_n,
\]
and $\mathbf{L}^{\star} = (L_{1}^{\star},\ldots, L_{p_n}^{\star})^T$,
where $\beta_0^M = \argmin_{\beta_0} E l(\beta_0, Y)$.
It can be shown that $EY = b'(\beta_0^M)$ and that $\overline{Y} =
b'(\hat{\beta}_0^M),$ where $\overline{Y}$ is the sample average.\vadjust{\goodbreak} We
sort the
vector $\mathbf{L}_n $ in a descent order and select a set of variables
\[
\widehat{\mathcal{N}}_{\nu_n} = \{1 \le j \le p_n\dvtx  L_{j,n} \ge
\nu_n \},
\]
where $\nu_n$ is a predefined threshold value. Such an
independence learning ranks the importance of features according
to their marginal contributions to the magnitudes of the
likelihood function. The marginal likelihood screening and the
MMLE screening share a common computation procedure as solving
$p_n$ optimization problems over a two-dimensional parameter
space. Hence the computation is much more feasible than
traditional variable selection methods.

Compared with MMLE screening, where the information utilized is
only the magnitudes of the estimators, the marginal likelihood
screening incorporates the whole contributions of the features to
the likelihood increments: both the magnitudes of the estimators
and their associated variation. Under the current condition (condition~C$'$), the variance of the MMLEs are at a comparable level (through
the magnitude of $V$, an implication of the convexity of the
objective functions), and the two screening methods are
equivalent. Otherwise, if $V$ depends on $n$, the marginal
likelihood screening can still preserve the nonsparsity
structure, while the MMLE screening may need some corresponding
adjustments, which we will not discuss in detail as it is beyond
the scope of the current paper.

Next we will show that the sure screening property holds under
certain conditions. Similarly to the MMLE screening, we first
build the theoretical foundation of the marginal likelihood
screening. That is, the marginal likelihood increment is also a
measurement of the correlation between the marginal covariate and
the mean response function.
%t6
\begin{theorem}\label{the-6}
For $j=1,\ldots, p_n$, the marginal likelihood increment
$L_j^{\star}=0$
if and only if $\operatorname{cov}(b'(\mathbf{X}^T
\bolds{\beta}^{\star}), X_j) = 0$.
\end{theorem}

As a direct corollary of Theorem~\ref{the-1}, we can easily show
the following corollary for the purpose of model selection
consistency.
%cor2
\begin{corollary}\label{cor-2}
If the partial orthogonality condition in Corollary~\ref{cor-1}
holds, then $L_j^{\star} =0$, for $j \notin
\mathcal{M}_{\star}.$
\end{corollary}

We can also strengthen the result of minimum signals as follows.
On the other hand, we also show that the total signals cannot be
too large. That is, there cannot be too many signals that exceed
certain threshold.
%t7
\begin{theorem}\label{the-7}
Under the conditions in Theorem~\ref{the-3} and the condition~C$'$,
we have
\[
\min_{j \in\mathcal{M}_{\star}}|L_j^{\star}|\ge c_6 n^{-2\kappa}\vadjust{\goodbreak}
\]
for some positive constant $c_6$, provided that $|\operatorname
{cov}(b'(\mathbf{X}^T
\bolds{\beta}^{\star}), X_j)| \ge c_1 n^{-\kappa}$ for $j \in
\mathcal{M}_{\star}$. If, in addition, conditions~F and G hold,
then
\[
\|\mathbf{L}^{\star}\| = O(\|\bolds{\beta}^M\|^2) = O(\|\bolds
{\Sigma}\bolds{\beta}^\star
\|^2) = O(\lambda_{\max}(\bolds{\Sigma})).
\]

\end{theorem}

The technical challenge is that the stochastic noise $\|\mathbf{L}_n -
\mathbf{L}^{\star}\|_{\infty}$ is usually of the order of $O(n^{-2
\kappa}
+ n^{-1/2} \log p_n )$, which can be an order of magnitude larger
than the signals given in Theorem~\ref{the-7}, unless $\kappa<
1/4$. Nevertheless, by a different trick that utilizes the fact
that ranking is invariant under a strict monotonic transform, we
are able to demonstrate the sure screening independence property for
$\kappa< 1/2$.
%t8
\begin{theorem}\label{the-8}
Suppose that conditions~\textup{A$'$, B$'$, C$'$} and \textup{D, E} and \textup{F} hold. Then, by
taking $\nu_n = c_7 n^{-2\kappa}$ for a sufficiently small $c_7
> 0$, there exists a $c_8 > 0$ such that
\begin{eqnarray*}
P (\mathcal{M}_{\star} \subset\widehat
{\mathcal{N}}_{\nu_n} ) \ge1 - s_n \bigl\{ \exp\bigl( - c_8
n^{1-2\kappa} /(k_nK_n)^2\bigr) + nm_1\exp(-m_0 K_n^\alpha)\bigr\}.
\end{eqnarray*}
\end{theorem}

Similarly to the MMLE screening, we can control the size of
$\widehat{\mathcal{N}}_{\nu_n}$ as follows. For simplicity of the
technical argument, we focus only on the case where $b''(\cdot)$
is bounded.
%t9
\begin{theorem}\label{the-9}
Under conditions~\textup{A$'$, B$'$, C$'$, D, F} and \textup{G}, if $b''(\cdot)$ is bounded,
then we have
\begin{eqnarray*}
& & P [ | \widehat{\mathcal{N}}_{\nu_n}| \leq
O\{n^{2\kappa}\lambda_{\max} (\bolds{\Sigma})\} ] \\
&&\qquad \geq 1 - p_n \bigl\{ \exp\bigl( - c_8 n^{1-2\kappa} /(k_nK_n)^2\bigr) +
nm_1\exp(-m_0 K_n^\alpha)\bigr\}.
\end{eqnarray*}
\end{theorem}

%s7 ###
\section{Numerical results}\label{sec7}
In this section, we present several simulation examples to evaluate the
performance of SIS procedure with generalized linear models. It was
demonstrated in Fan and Lv (\citeyear{FanLv08}) and \citet{Fan08} that
independent screening is a fast but crude method of reducing the
dimensionality to a more moderate size. Some methodological extensions
include iterative SIS (ISIS) and multi-stage procedures, such as
SIS-SCAD and SIS-LASSO, can be applied to perform the final variable
selection and parameter estimation simultaneously. Extensive
simulations on these procedures were also presented in \citet{Fan08}. To avoid repetition, in this paper, we focus on the vanilla
SIS, and aim to evaluate the sure screening property and to demonstrate
some factors influencing the false selection rate. We vary the sample
size from $80$ to $600$ for different scenarios to gauge the
difficulties of the simulation models. The following three
configurations with {$p=2000$, $5000$ and $40{,}000$} predictor variables
are considered for generating the covariates $\mathbf{X}=(X_1,\ldots,X_p)^T$:
\begin{enumerate}[S3.]
\item[S1.]
The covariates are generated according to
%e5 ###
%
\begin{equation}\label{revise-e1}
X_j = \frac{\varepsilon_j + a_j \varepsilon}{\sqrt{1 + a_j^2}},
\end{equation}
where\vspace*{-2pt} $\varepsilon$ and $\{\varepsilon_j\}_{j=1}^{[p/3]}$ are i.i.d. standard
normal random variables,\break
$\{\varepsilon_j\}_{j=[p/3]+1}^{[2p/3]}$ are i.i.d. and follow a double
exponential distributions with location parameter zero and scale
parameter one and $\{\varepsilon_j\}_{j=[2p/3]+1}^{p}$ are i.i.d. and follow
a mixture normal distribution with two components $N(-1,1)$, $N(1,0.5)$
and equal mixture proportion. The covariates are standardized to be
mean zero and variance one.
The constants $\{a_j\}_{j=1}^{q}$ are the same\vspace*{1pt} and chosen such that the
correlation
$\rho=\operatorname{corr}(X_i, X_j)=0, 0.2,0.4,0.6$ and $0.8$, among the
first $q$ variables, and $a_j =0$ for $j>q$. The parameter $q$ is also
related to the overall correlation in the covariance matrix. We will
present the numerical results with $q=15$ for this setting.

\item[S2.]
The covariates are also generated from (\ref{revise-e1}), except that
$\{a_j\}_{j=1}^{q}$ are i.i.d. normal random variables with mean $a$
and variance $1$ and $a_j = 0$ for $j > q$. The value of $a$ is taken
such that $E\operatorname{corr}(X_i, X_j) =0,0.2,0.4,0.6$ and $0.8$, among
the first $q$ variables. The simulation results to be presented for
this setting use $q=50$.

\item[S3.]
Let $\{X_j\}_{j=1}^{p-50}$ be i.i.d. standard normal random variables and
\[
X_k = \sum_{j=1}^s X_j (-1)^{j+1}/5 + \sqrt{25-s}/5\varepsilon_k, \qquad k=p-49,\ldots,p,
\]
where $\{\varepsilon_k\}_{k=p-49}^{p}$ are standard normally distributed.
\end{enumerate}

%t1 ###
%
\begin{table}
\tabcolsep=0pt
\caption{The median of the 200 empirical maximum eigenvalues, with its
robust estimate of~SD~in~the~parenthesis, of the corresponding sample
covariance matrices of~covariates~based~200 simulations with partial
combinations of $n=80, 300, 600$, $p=2000, 5000, 40{,}000$ and $q=15, 50$
in the first two settings (\textup{S1} and \textup{S2})}\label{table-1}
\begin{tabular*}{\textwidth}{@{\extracolsep{4in minus 4in}}lc d{3.8}d{3.8}d{3.7}d{3.7}d{3.7}@{\hspace*{-2pt}}}
\hline
& & \multicolumn{5}{c@{}}{$\bolds{\rho}$} \\[-6pt]
& & \multicolumn{5}{c@{}}{\hrulefill} \\
$\bolds{(p,n)}$ & \textbf{Setting} & \textbf{0} & \textbf{0.2} & \textbf{0.4} & \textbf{0.6} & \textbf{0.8} \\\hline
(40,000, 80) & S1 ($q=15$)                           & 549.9\ (1.4) & 550.1\ (1.4) & 550.1\ (1.3) &550.1\ (1.3) & 550.1\ (1.4) \\
(40,000, 80) & S2 ($q=50$)                           & 550.0\ (1.4) & 550.1\ (1.4) & 550.4\ (1.5) &552.9\ (1.8) & 558.5\ (2.4) \\
(40,000, 300) & S1 ($q=15$)                          & 157.3\ (0.4) & 157.4\ (0.4) & 157.4\ (0.4) &157.4\ (0.3) & 157.7\ (0.4)\\
(40,000, 300) & S2 ($q=50$)                          & 157.4\ (0.4) & 157.5\ (0.4) & 160.9\ (1.2) &168.2\ (1.0) & 176.9\ (1.0) \\
(5000, 300) & S1 ($q=15$)                & 25.68\ (0.2) & 25.68\ (0.2) & 26.18\ (0.2) &27.99\ (0.4) & 30.28\ (0.4) \\
(5000, 300) & S1 ($q=50$)                & 25.69\ (0.1) & 29.06\ (0.5) &37.98\ (0.7) & 47.49\ (0.7) & 57.17\ (0.5) \\
(2000, 600) & S1 ($q=15$)                & 7.92\ (0.07) & 8.32\ (0.15) & 10.5\ (0.3) &13.09\ (0.3)& 15.79\ (0.2) \\
(2000, 600) & S1 ($q=50$)                & 7.93\ (0.07) & 14.62\ (0.40) &23.95\ (0.7)&33.90\ (0.6) &43.56\ (0.5) \\
(2000, 600) & S2 ($q=50$)                & 7.93\ (0.07) & 14.62\ (0.40) & 23.95\ (0.7)&33.90\ (0.6) &43.56\ (0.5) \\
\hline
\end{tabular*}
\end{table}

Table~\ref{table-1} summarizes the median of the empirical maximum eigenvalues of
the covariance matrix and its robust estimate of the standard deviation
(RSD) based 200 simulations in the first two settings (S1 and S2) with
partial combinations of sample size $n=80, 300, 600$, $p=2000, 5000,
40{,}000$ and $q=15, 50$. RSD is the interquantile range (IQR) divided by
1.34. The empirical maximum eigenvalues are always larger than their
population version, depending on the realizations of the design matrix.
The empirical minimum eigenvalue is always zero, and the empirical
condition numbers for the sample covariance matrix are infinite, since
$p > n$. Generally, the empirical maximum eigenvalues increase as the
correlation parameters $\rho$, $q$, the numbers of covariates $p$
increase, and/or the sample sizes $n$ decrease.

With these three settings, we aim to illustrate the behaviors of the
two SIS procedures under different correlation structures. For each
simulation and each model, we apply the two SIS procedures, the
marginal MLE and the marginal likelihood ratio methods, to screen
variables. The minimum model size (MMS)\vadjust{\goodbreak} required for each method to
have a sure screening, that is, to contain the true model $\mathcal
{M}_{\star}$, is used as a measure of the effectiveness of a screening
method. This avoids the issues of choosing the thresholding parameter.
To gauge the difficulty of the problem, we also include the LASSO and
the SCAD as references for comparison when $p=2000$ and $5000$. The
smaller $p$ is used due to the computation burden of the LASSO and the
SCAD. In addition, as demonstrated in our simulation results, they do
not perform well when $p$ is large. Our initial intension is to
demonstrate that the simple SIS does not perform much worse than the
far more complicated procedures like the LASSO and the SCAD. To our
surprise, the SIS can even outperform those more complicated methods in
terms of variable screening. Again, we record the MMS for the LASSO and
the SCAD for each simulation and each model, which does not depend on
the choice of regularization parameters. When the LASSO or the SCAD cannot recover the true model even with the smallest regularization
parameter, we average the model size with the smallest regularization
parameter and $p$. These interpolated MMS' are presented with italic
font in Tables~\ref{table-3}--\ref{table-5} and~\ref{table-9} to distinguish from the real MMS. Results for
logistic regressions and linear regressions are presented in the
following two subsections.

%s7.1 ###
\subsection{Logistic regressions}
The generated data $(\mathbf{X}_1^T, Y_1), \ldots, (\mathbf{X}_n^T,
Y_n)$ are $n$
i.i.d. copies of a pair $(\mathbf{X}^T, Y)$, in which the conditional
distribution of the response $Y$ given $\mathbf{X}= \mathbf{x}$ is binomial
distribution with probability of success $p(\mathbf{x}) = \exp
(\mathbf{x}^T\bolds{\beta}^{\star})/[1+\exp(\mathbf{x}^T\bolds
{\beta}^{\star})]$.
We vary the size of the nonsparse set of coefficients as
$s=3,6,12,15$ and $24$. For each simulation, we evaluate each method
by summarizing the median minimum model size (MMMS) of the selected
models as well as its associated RSD, which is the associated
interquartile range (IQR) divided by 1.34. The results, based on 200
simulations for each scenario are recorded in the second and third
panel of Table~\ref{table-2} and the second panel of Tables~\ref{table-3}--\ref{table-5}. Specifically,
Table~\ref{table-2} records the MMMS and the associated RSD for SIS under the first
two settings when $p=40{,}000$, while Tables~\ref{table-3}--\ref{table-5} record these results for
SIS, the LASSO and the SCAD when $p=2000$ and $5000$ under Settings 1, 2 and 3, respectively. The true parameters are also recorded in each
corresponding table.

%t2 ###
%
\begin{table}
\tabcolsep=7pt
\caption{The MMMS and the associated RSD (in the parenthesis) of the
simulated examples for~logistic regressions in the first two settings
(\textup{S1} and \textup{S2}) when $p=40{,}000$}\label{table-2}
\begin{tabular*}{\textwidth}{@{\extracolsep{4in minus 4in}}lc e{3.7}e{3.7}c e{3.7}e{3.7}@{\hspace*{-2pt}}}
\hline
$\bolds{\rho}$&$\bolds{n}$&
\multicolumn{1}{c}{\textbf{SIS-MLR}} & \multicolumn{1}{c}{\textbf{SIS-MMLE}} &$\bolds{n}$&
\multicolumn{1}{c}{\textbf{SIS-MLR}} & \multicolumn{1}{c@{}}{\textbf{SIS-MMLE}} \\
\hline
Setting 1, $q=15$\\
%&&&&&&&&\\
&\multicolumn{3}{c}{$s=3$, $\bolds{\beta}^{\star}=(1,1.3,1)^T$} &
\multicolumn{3}{c@{}} {$s=6$, $\bolds{\beta}^{\star}=(1,1,3,1,\ldots
)^T$} \\
\quad 0 & 300 & 87,.5\ (381) & 89,\ (375) & 300 & 47,\ (164) & 50,\ (170) \\
\quad 0.2 & 200 & 3,\ (0) & 3,\ (0) & 300 & 6,\ (0) & 6,\ (0) \\
\quad 0.4 & 200 & 3,\ (0) & 3,\ (0) & 300 & 7,\ (1) & 7,\ (1) \\
\quad 0.6 & 200 & 3,\ (1) & 3,\ (1) & 300 & 8,\ (1) & 8,\ (2) \\
\quad 0.8 & 200 & 4,\ (1) & 4,\ (1) & 300 & 9,\ (3) & 9,\ (3) \\[6pt]
&\multicolumn{3}{c}{$s=12$, $\bolds{\beta}^{\star}=(1,1.3,\ldots)^T$} &
\multicolumn{3}{c@{}}{$s=15$, $\bolds{\beta}^{\star}=(1,1.3,\ldots)^T$}\\
\quad 0 & 500& 297,\ (589) & 302,.5\ (597) & 600& 350,\ (607) & 359,.5\ (612) \\
\quad 0.2 & 300& 13,\ (1) & 13,\ (1) & 300& 15,\ (0) & 15,\ (0) \\
\quad 0.4 & 300& 14,\ (1) & 14,\ (1) & 300& 15,\ (0) & 15,\ (0) \\
\quad 0.6 & 300& 14,\ (1) & 14,\ (1) & 300& 15,\ (0) & 15,\ (0) \\
\quad 0.8 & 300& 14,\ (1) & 14,\ (1) & 300& 15,\ (0) & 15,\ (0) \\
[6pt]
Setting 2, $q=50$ \\
&\multicolumn{3}{c}{$s=3$, $\bolds{\beta}^{\star}=(1,1.3,1)^T$} &
\multicolumn{3}{c@{}} {$s=6$, $\bolds{\beta}^{\star}=(1,1.3,1,\ldots)^T$} \\
\quad 0 & 300 & 84,.5\ (376) & 88,.5\ (383) & 500 & 6,\ (1) & 6,\ (1) \\
\quad 0.2 & 300 & 3,\ (0) & 3,\ (0) & 500 & 6,\ (0) & 6,\ (0) \\
\quad 0.4 & 300 & 3,\ (0) & 3,\ (0) & 500 & 6,\ (1) & 6,\ (1) \\
\quad 0.6 & 300 & 3,\ (1) & 3,\ (1) & 500 & 8,.5\ (4) & 9,\ (5) \\
\quad 0.8 & 300 & 5,\ (4) & 5,\ (4) & 500 & 13,.5\ (8) & 14,\ (8) \\[6pt]
&\multicolumn{3}{c}{$s=12$, $\bolds{\beta}^{\star}=(1,1.3,\ldots)^T$} &
\multicolumn{3}{c@{}}{$s=15$, $\bolds{\beta}^{\star}=(1,1.3,\ldots
)^T$}\\
\quad 0 &600& 77,\ (114) & 78,.5\ (118) &800& 46,\ (82) & 47,\ (83) \\
\quad 0.2 &500& 18,\ (7) & 18,\ (7) &500& 26,\ (6) & 26,\ (6) \\
\quad 0.4 &500& 25,\ (8) & 25,\ (10) &500& 34,\ (7) & 33,\ (8) \\
\quad 0.6 &500& 32,\ (9) & 31,\ (8) &500& 39,\ (7) & 38,\ (7) \\
\quad 0.8 &500& 36,\ (8) & 35,\ (9) &500& 40,\ (6) & 42,\ (7) \\
\hline
\end{tabular*}
\end{table}

\begin{figure}

\includegraphics{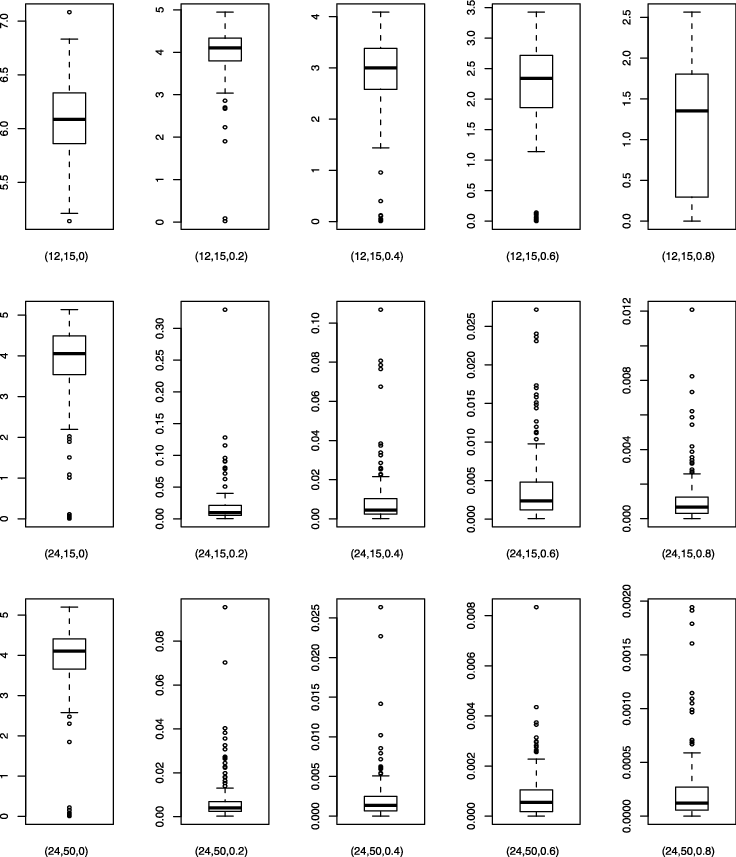}

\caption{The boxplots of the minimum $|t|$-statistics in the oracle
models among 200 simulations for the first setting (\textup{S1}) with logistic
regression examples with $\bolds{\beta}^{\star}=(3,4,\ldots)^T$ when
$s=12,24$, $q=15, 50$, $n=600$ and $p=2000.$ The triplets under each
plot represent the corresponding values of $(s, q, \rho)$, respectively.}\label{f1}
\end{figure}

To demonstrate the difficulty of our simulated models, we depict the
distribution, among 200 simulations, of the minimum $|t|$-statistics of
$s$ estimated regression coefficients in the oracle model in which the
statistician does not know that all variables are statistically
significant. This shows the difficulty in recovering all significant
variables even in the oracle model with the minimum model size $s$. The
distribution was computed for each setting and scenario but only a few
selected settings are shown presented in Figure~\ref{f1}. In fact, the
distributions under Setting 1 are very similar to those under Setting 2
when the same $q$ value is taken. It can be seen that the magnitude of
the minimum $|t|$-statistics is reasonably small and getting smaller as
the correlation within covariates (measured by $\rho$ and $q$)
increases, sometimes achieving three decimals. Given such small
signal-to-noise ratio in the oracle models, the difficulty of our
simulation models is a self-evident even if the signals seem not that small.

%t3 ###
%
\begin{table}
\tabcolsep=0pt
\caption{The MMMS and the associated RSD (in the parenthesis) of the
simulated examples for~logistic regressions in Setting 1 (\textup{S1}) when
$p=5000$ and $q=15$. The values with~italic~font indicate that the LASSO or
the SCAD cannot recover the true model~even~with~smallest
regularization parameter and are estimated}\label{table-3}
\begin{tabular*}{\textwidth}{@{\extracolsep{4in minus 4in}}lc e{3.7}e{3.7}e{3.8}e{3.7}@{\hspace*{-2pt}}}
\hline
$\bolds{\rho}$&$\bolds{n}$&
\multicolumn{1}{c}{\textbf{SIS-MLR}} & \multicolumn{1}{c}{\textbf{SIS-MMLE}}& \multicolumn{1}{c}{\textbf{LASSO}}&
\multicolumn{1}{c@{}}{\textbf{SCAD}}\\
\hline
% \multicolumn{6}{c}{Setting 1, p=5000, q=15} \\
$s=3$, $\bolds{\beta}^{\star}=(1,1.3,1)^T$ \\
\quad 0 & 300 & 3,\ (0) & 3,\ (0) & 3,\ (1) & 3,\ (1) \\
\quad 0.2 & 300 & 3,\ (0) & 3,\ (0) & 3,\ (0) & 3,\ (0) \\
\quad 0.4 & 300 & 3,\ (0) & 3,\ (0) & 3,\ (0) & 3,\ (0) \\
\quad 0.6 & 300 & 3,\ (0) & 3,\ (0) & 3,\ (0) & 3,\ (1) \\
\quad 0.8 & 300 & 3,\ (1) & 3,\ (1) & 4,\ (1) & 4,\ (1) \\[6pt]
$s=6$, $\bolds{\beta}^{\star}=(1,1.3,1,1.3,1,1.3)^T$ \\
\quad 0 & 200 & 8,\ (6) & 9,\ (7) & 7,\ (1) & 7,\ (1) \\
\quad 0.2 & 200 & 18,\ (38) & 20,\ (39) & 9,\ (4) & 9,\ (2) \\
\quad 0.4 & 200 & 51,\ (77) & 64,.5\ (76) & 20,\ (10) & 16,.5\ (6) \\
\quad 0.6 & 300 & 77,.5\ (139) & 77,.5\ (132) & 20,\ (13) & 19,\ (9) \\
\quad 0.8 & 400 & 306,.5\ (347) & 313,\ (336) & 86,\ (40) & 70,.5\ (35) \\[6pt]
$s=12$, $\bolds{\beta}^{\star}=(1,1.3,\ldots)^T$ \\
\quad 0 & 300 & 297,.5\ (359) & 300,\ (361) & 72,.5\ (3704) & 12,\ (0) \\
\quad 0.2 & 300 & 13,\ (1) & 13,\ (1) & 12,\ (1) & 12,\ (0) \\
\quad 0.4 & 300 & 14,\ (1) & 14,\ (1) & 14,\ (1861) & 13,\ (1865) \\
\quad 0.6 & 300 & 14,\ (1) & 14,\ (1) & \mathit{2552},\ \mathit{(85)} & 12,\ (3721) \\
\quad 0.8 & 300 & 14,\ (1) & 14,\ (1)& \mathit{2556},\ \mathit{(10)} & 12,\ (3722) \\[6pt]
$s=15$, $\bolds{\beta}^{\star}=(3,4,\ldots)^T$\\
\quad 0 & 300 & 479,\ (622) & 482,\ (615) & 69,.5\ (68) & 15,\ (0) \\
\quad 0.2 & 300 & 15,\ (0) & 15,\ (0) & 16,\ (13) & 15,\ (0) \\
\quad 0.4 & 300 & 15,\ (0) & 15,\ (0) & 38,\ (3719) & 15,\ (3720) \\
\quad 0.6 & 300 & 15,\ (0) & 15,\ (0) & \mathit{2555},\ \mathit{(87)} & 15,\ (1472) \\
\quad 0.8 & 300 & 15,\ (0) & 15,\ (0) & \mathit{2552},\ \mathit{(8)} & 15,\ (1322) \\
\hline
\end{tabular*}\vspace*{5pt}
\end{table}

%t4 ###
%
\begin{table}
\caption{The MMMS and the associated RSD (in the parenthesis) of the
simulated examples for~logistic regressions in Setting 2 (\textup{S2}) when
$p=2000$ and $q=50$. The values with~italic~font have the same meaning as Table~\protect\ref{table-2}}\label{table-4}
\begin{tabular*}{\textwidth}{@{\extracolsep{4in minus 4in}}lc e{3.8}e{3.8}e{4.4}e{4.6}@{}}
\hline
$\bolds{\rho}$&$\bolds{n}$&
\multicolumn{1}{c}{\textbf{SIS-MLR}} & \multicolumn{1}{c}{\textbf{SIS-MMLE}}& \multicolumn{1}{c}{\textbf{LASSO}}&
\multicolumn{1}{c@{}}{\textbf{SCAD}}\\
\hline
% \multicolumn{6}{c}{Setting 2, p=2000, q=50} \\
\multicolumn{6}{@{}l@{}}{$s=3$, $\bolds{\beta}^{\star}=(3,4,3)^T$} \\
\quad 0 & 200 & 3,\ (0) & 3,\ (0) & 3,\ (0) & 3,\ (0) \\
\quad 0.2 & 200 & 3,\ (0) & 3,\ (0) & 3,\ (0) & 3,\ (0) \\
\quad 0.4 & 200 & 3,\ (0) & 3,\ (0) & 3,\ (0) & 3,\ (1) \\
\quad 0.6 & 200 & 3,\ (1) & 3,\ (1) & 3,\ (1) & 3,\ (1) \\
\quad 0.8 & 200 & 5,\ (5) & 5,.5\ (5) & 6,\ (4) & 6,\ (4) \\[6pt]
\multicolumn{6}{@{}l@{}}{$s=6$, $\bolds{\beta}^{\star}=(3,-3,3,-3,3,-3)^T$} \\
\quad 0 & 200 & 8,\ (6) & 9,\ (7) & 7,\ (1) & 7,\ (1) \\
\quad 0.2 & 200 & 18,\ (38) & 20,\ (39) & 9,\ (4) & 9,\ (2) \\
\quad 0.4 & 200 & 51,\ (77) & 64,.5\ (76) & 20,\ (10) & 16,.5\ (6) \\
\quad 0.6 & 300 & 77,.5\ (139) & 77,.5\ (132) & 20,\ (13) & 19,\ (9) \\
\quad 0.8 & 400 & 306,.5\ (347) & 313,\ (336) & 86,\ (40) & 70,.5\ (35) \\[6pt]
\multicolumn{6}{@{}l@{}}{$s=12$, $\bolds{\beta}^{\star}=(3,4,\ldots)^T$} \\
\quad 0 &600& 13,\ (6) & 13,\ (7) & 12,\ (0) & 12,\ (0) \\
\quad 0.2 &600& 19,\ (6) & 19,\ (6) & 13,\ (1) & 13,\ (2) \\
\quad 0.4 &600& 32,\ (10) & 30,\ (10)& 18,\ (3) & 17,\ (4) \\
\quad 0.6 &600& 38,\ (9) & 38,\ (10) & 22,\ (3) & 22,\ (4) \\
\quad 0.8 &600& 38,\ (7) & 39,\ (8) & \mathit{1071},\ \mathit{(6)} & \mathit{1042},\ \mathit{(34)} \\[6pt]
\multicolumn{6}{@{}l@{}}{$s=24$, $\bolds{\beta}^{\star}=(3,4,\ldots)^T$}\\
\quad 0 &600& 180,\ (240) & 182,\ (238) & 35,\ (9) & 31,\ (10) \\
\quad 0.2 &600& 45,\ (4) & 45,\ (4) & 35,\ (27) & 32,\ (24) \\
\quad 0.4 &600& 46,\ (3) & 47,\ (2) & \mathit{1099},\ \mathit{(17)} & 1093,\ (1456) \\
\quad 0.6 &600& 48,\ (2) & 48,\ (2) & \mathit{1078},\ \mathit{(5)} & \mathit{1065},\ \mathit{(23)} \\
\quad 0.8 &600& 48,\ (1) & 48,\ (1) & \mathit{1072},\ \mathit{(4)} & \textit{1067},\ \mathit{(13)} \\
\hline
\end{tabular*}
\end{table}

%t5 ###
%
\begin{table}[b!]
\caption{The MMMS and the associated RSD (in the parenthesis) of the
simulated examples~for~logistic regressions in Setting 3 \textup{(S3)} when
$p=2000$ and $n=600$. The~values~with~italic~font have the same meaning as
Table~\protect\ref{table-2}. M-$\lambda_{\max}$ and its~RSD~have~the~same~meaning as
Table~\protect\ref{table-1}}\label{table-5}
\begin{tabular*}{\textwidth}{@{\extracolsep{4in minus 4in}}le{2.8}e{3.6}e{3.6}e{3.4}e{3.4}@{}}
\hline
$\bolds{s}$ & \multicolumn{1}{c}{$\bolds{M}$\textbf{-}$\bolds{\lambda_{\max}}$(RSD)}&
\multicolumn{1}{c}{\textbf{SIS-MLR}} & \multicolumn{1}{c}{\textbf{SIS-MMLE}}& \multicolumn{1}{c}{\textbf{LASSO}}&
\multicolumn{1}{c@{}}{\textbf{SCAD}}\\
\hline
\phantom{0}3 & 8,.47\ (0.17)& 3,\ (0) & 3,\ (0) & 3,\ (1) & 3,\ (0) \\
\phantom{0}6 & 10,.36\ (0.26)& 56,\ (0) & 56,\ (0) & \mathit{1227},\ \mathit{(7)} & \mathit{1142},\ \mathit{(64)}\\
12&14,.69\ (0.39)& 63,\ (6) & 63,\ (6) & \mathit{1148},\ \mathit{(8)} & \mathit{1093},\ \mathit{(59)}\\
24&23,.70\ (0.14)& 214,.5\ (93) & 208,.5\ (82) & \mathit{1120},\ \mathit{(5)} & \mathit{1087},\ \mathit{(24)} \\
\hline
\end{tabular*}
\end{table}

%f1 ###
%

The MMMS and RSD with fixed correlation (S1) and random correlation
(S2) are comparable under the same $q$. As the correlation increases
and/or the nonsparse set size increases, the MMMS and the associated
RSD usually increase for all SIS, the LASSO and the SCAD. Among all the
designed scenarios of Settings 1 and~2, SIS performs well, while the
LASSO and the SCAD occasionally fail under very high correlations and
relatively large nonsparse set size ($s=12$, 15 and 24). Interestingly,
correlation within covariates can sometimes help SIS reduce the false
selection rate, as it can increase the marginal signals. It is notable
that the LASSO and the SCAD usually cannot select the important
variables in the third setting, due to the violation of the
irrepresentable condition for $s=6, 12$ and $24$, while SIS perform
reasonably well.

%s7.2 ###
\subsection{Linear models}
The generated data $(\mathbf{X}_1^T, Y_1), \ldots, (\mathbf{X}_n^T,
Y_n)$ are $n$
i.i.d. copies of a pair $(\mathbf{X}^T, Y)$, in which the response $Y$
follows a linear model with $Y = \mathbf{X}^T \bolds{\beta}^{\star}
+ \varepsilon,$
where the random error $\varepsilon$ is standard normally distributed. The
covariates are generated in the same manner as the logistic regression
settings. We take the same true coefficients and correlation structures
for part of the scenarios ($p=40{,}000$) as the logistic regression
examples, while vary the true coefficients for other scenarios, to
gauge the difficulty of the problem. The sample size for each scenario
is correspondingly decreased to reflect the fact that the linear model
is more informative. The results are recorded in
Tables~\ref{table-6}--\ref{table-9},
respectively. The trend of the MMMS and the associated RSD of SIS, the
LASSO and the SCAD varying with the correlation and/or the nonsparse
set size are similar to these
in the logistic regression examples, but
their magnitudes are usually smaller in the linear regression examples,
as the model is more informative. Overall, the SIS does a very
reasonable job in screening irrelevant variables and sometimes
outperforms the LASSO and the SCAD.

%t6 ###
%
\begin{table}
\caption{The MMMS and the associated RSD (in the parenthesis) of the
simulated examples in~the~first two settings (\textup{S1} and \textup{S2}) for linear
regressions when $p=40{,}000$}\label{table-6}
\begin{tabular*}{\textwidth}{@{\extracolsep{4in minus 4in}}ld{3.0}e{2.5}e{3.5}c e{3.7}e{3.7}@{\hspace*{-2pt}}}
\hline
$\bolds{\rho}$
& \multicolumn{1}{c}{$\bolds{n}$}
& \multicolumn{1}{c}{\textbf{SIS-MLR}}
& \multicolumn{1}{c}{\textbf{SIS-MMLE}}
& \multicolumn{1}{c}{$\bolds{n}$}
& \multicolumn{1}{c}{\textbf{SIS-MLR}}
& \multicolumn{1}{c@{}}{\textbf{SIS-MMLE}} \\
\hline
\multicolumn{7}{@{}l@{}}{Setting 1, $q=15$} \\
%&&&&&&&&\\
&\multicolumn{3}{c}{$s=3$, $\bolds{\beta}^{\star}=(1,1.3,1)^T$} &
\multicolumn{3}{c@{}}{$s=6$, $\bolds{\beta}^{\star}=(1,1,3,1,\ldots)^T$} \\
\quad 0 & 80 & 12,\ (18) & 12,\ (18) & 150 & 42,\ (157) & 42,\ (157) \\
\quad 0.2 & 80 & 3,\ (0) & 3,\ (0) & 150 & 6,\ (0) & 6,\ (0) \\
\quad 0.4 & 80 & 3,\ (0) & 3,\ (0) & 150 & 6,.5\ (1) & 6,.5\ (1) \\
\quad 0.6 & 80 & 3,\ (0) & 3,\ (0) & 150 & 6,\ (1) & 6,\ (1) \\
\quad 0.8 & 80 & 3,\ (0) & 3,\ (0) & 150 & 7,\ (1) & 7,\ (1)
\\[6pt]
&\multicolumn{3}{c}{$s=12$, $\bolds{\beta}^{\star}=(1,1.3,\ldots)^T$} &
\multicolumn{3}{c@{}}{$s=15$, $\bolds{\beta}^{\star}=(1,1.3,\ldots)^T$}\\
\quad 0 & 300& 143,\ (282)& 143,\ (282) & 400& 135,.5\ (167) & 135,.5\ (167) \\
\quad 0.2 & 200& 13,\ (1) & 13,\ (1) & 200& 15,\ (0) & 15,\ (0) \\
\quad 0.4 & 200& 13,\ (1) & 13,\ (1) & 200& 15,\ (0) & 15,\ (0) \\
\quad 0.6 & 200& 13,\ (1) & 13,\ (1) & 200& 15,\ (0) & 15,\ (0) \\
\quad 0.8 & 200& 13,\ (1) & 13,\ (1) & 200& 15,\ (0) & 15,\ (0)
\\[6pt]
\multicolumn{7}{@{}l@{}}{Setting 2, $q=50$} \\
&\multicolumn{3}{c}{$s=3$, $\bolds{\beta}^{\star}=(1,1.3,1)^T$} &
\multicolumn{3}{c@{}}{$s=6$, $\bolds{\beta}^{\star}=(1,1.3,1,\ldots)^T$} \\
\quad 0 & 100 & 3,\ (2) & 3,\ (2) & 200 & 7,.5\ (7) & 7,.5\ (7) \\
\quad 0.2 & 100 & 3,\ (0) & 3,\ (0) & 200 & 6,\ (1) & 6,\ (1) \\
\quad 0.4 & 100 & 3,\ (0) & 3,\ (0) & 200 & 7,\ (1) & 7,\ (1) \\
\quad 0.6 & 100 & 3,\ (0) & 3,\ (0) & 200 & 7,\ (2) & 7,\ (2) \\
\quad 0.8 & 100 & 3,\ (1) & 3,\ (1) & 200 & 8,\ (4) & 8,\ (4)
\\[6pt]
&\multicolumn{3}{c}{$s=12$, $\bolds{\beta}^{\star}=(1,1.3,\ldots)^T$} &
\multicolumn{3}{c@{}}{$s=15$, $\bolds{\beta}^{\star}=(1,1.3,\ldots)^T$}\\
\quad 0 & 400& 22,\ (27) & 22,\ (27) &500& 35,\ (52) & 35,\ (52) \\
\quad 0.2 &300& 16,\ (5) & 16,\ (5) &300& 24,\ (7) & 24,\ (7) \\
\quad 0.4 &300& 19,\ (8) & 19,\ (8) &300& 30,\ (10) & 30,\ (10) \\
\quad 0.6 &300& 25,\ (8) & 25,\ (8) &300& 33,.5\ (7) & 33,.5\ (7) \\
\quad 0.8 &300& 24,\ (7) & 24,\ (7) &300& 35,\ (8) & 35,\ (8) \\
\hline
\end{tabular*}\vspace*{-3pt}
\end{table}

%t7 ###
%
\begin{table}
\caption{The MMMS and the RSD (in the parenthesis) of the simulated
examples for linear regressions in Setting 1 (\textup{S1}) when $p=5000$ and $q=15$}\label{table-7}
\begin{tabular*}{\textwidth}{@{\extracolsep{4in minus 4in}}lc e{3.7}e{3.7}e{3.7}e{2.4}@{}}
\hline
$\bolds{\rho}$& $\bolds{n}$
& \multicolumn{1}{c}{\textbf{SIS-MLR}}
& \multicolumn{1}{c}{\textbf{SIS-MMLE}}
& \multicolumn{1}{c}{\textbf{LASSO}}
& \multicolumn{1}{c@{}}{\textbf{SCAD}}\\
\hline
% \multicolumn{6}{c}{Setting 1, p=5000, q=15} \\
\multicolumn{6}{@{}l@{}}{$s=3$, $\bolds{\beta}^{\star}=(0.5,0.67,0.5)^T$} \\
\quad 0 & 100 & 12,\ (40) & 12,\ (40) & 3,\ (1) & 3,\ (1) \\
\quad 0.2 & 100 & 3,\ (1) & 3,\ (1) & 3,\ (0) & 3,\ (0) \\
\quad 0.4 & 100 & 3,\ (0) & 3,\ (0) & 3,\ (0) & 3,\ (0) \\
\quad 0.6 & 100 & 3,\ (1) & 3,\ (1) & 5,\ (7) & 5,\ (5) \\
\quad 0.8 & 100 & 4,\ (2) & 4,\ (2) & 4,\ (1) & 4,\ (1) \\[5pt]
\multicolumn{6}{@{}l@{}}{$s=6$, $\bolds{\beta}^{\star}=(0.5,0.67,0.5,0.67,0.5,0.67)^T$} \\
\quad 0 & 100 & 210,.5\ (422) & 210,.5\ (422) & 33,.5\ (651) & 25,\ (22) \\
\quad 0.2 & 100 & 7,\ (2) & 7,\ (2) & 6,\ (1) & 6,\ (1) \\
\quad 0.4 & 100 & 7,\ (2) & 7,\ (2) & 6,\ (1) & 6,\ (1) \\
\quad 0.6 & 100 & 8,\ (2) & 8,\ (2) & 7,\ (1) & 7,\ (1) \\
\quad 0.8 & 100 & 9,\ (3) & 9,\ (3) & 7,\ (2) & 8,\ (1) \\[5pt]
\multicolumn{6}{@{}l@{}}{$s=12$, $\bolds{\beta}^{\star}=(0.5,0.67,\ldots)^T$} \\
\quad 0 & 300 & 49,\ (76) & 49,\ (76) & 12,\ (1) & 12,\ (0) \\
\quad 0.2 & 100 & 14,\ (2) & 14,\ (2) & 12,\ (1) & 12,\ (1) \\
\quad 0.4 & 100 & 14,\ (1) & 14,\ (1) & 12,\ (1) & 12,\ (1) \\
\quad 0.6 & 100 & 14,\ (1) & 14,\ (1) & 13,\ (1) & 13,\ (1) \\
\quad 0.8 & 100 & 14,\ (1) & 14,\ (1) & 13,\ (1) & 13,\ (1) \\[5pt]
\multicolumn{6}{@{}l@{}}{$s=15$, $\bolds{\beta}^{\star}=(0.5,0.67,\ldots)^T$}\\
\quad 0 & 300 & 199,\ (251) & 199,\ (251) & 17,\ (2) & 15,\ (0) \\
\quad 0.2 & 100 & 17,\ (5) & 17,\ (5) & 15,\ (1) & 15,\ (0) \\
\quad 0.4 & 100 & 15,\ (0) & 15,\ (0) & 15,\ (0) & 15,\ (0) \\
\quad 0.6 & 100 & 15,\ (0) & 15,\ (0) & 15,\ (0) & 15,\ (0) \\
\quad 0.8 & 100 & 15,\ (0) & 15,\ (0) & 15,\ (0) & 15,\ (1) \\
\hline
\end{tabular*}   \vspace*{-3pt}
\end{table}

%t8 ###
%
\begin{table}
\caption{The MMMS and the RSD (in the parenthesis) of the simulated
examples for~linear~regressions in Setting 2 (\textup{S2}) when $p=2000$ and $q=50$}\label{table-8}
\begin{tabular*}{\textwidth}{@{\extracolsep{4in minus 4in}}lc e{3.6}e{3.7}e{3.6}e{2.4}@{}}
\hline
$\bolds{\rho}$& $\bolds{n}$&
\multicolumn{1}{c}{\textbf{SIS-MLR}} & \multicolumn{1}{c}{\textbf{SIS-MMLE}}& \multicolumn{1}{c}{\textbf{LASSO}}& \multicolumn{1}{c@{}}{\textbf{SCAD}}\\
\hline
% \multicolumn{6}{c}{Setting 2, p=2000, q=50} \\
\multicolumn{6}{@{}l@{}}{$s=3$, $\bolds{\beta}^{\star}=(0.6,0.8,0.6)^T$}\\
\quad 0 & 100 & 5,\ (14) & 6,\ (16) & 4,\ (4) & 4,\ (2) \\
\quad 0.2 & 100 & 3,\ (1) & 3,\ (1) & 3,\ (0) & 3,\ (0) \\
\quad 0.4 & 100 & 3,\ (1) & 4,\ (1) & 3,\ (1) & 3,\ (1) \\
\quad 0.6 & 100 & 5,\ (3) & 7,\ (5) & 4,\ (1) & 4,\ (1) \\
\quad 0.8 & 100 & 7,\ (7) & 14,\ (12) & 5,\ (57) & 7,\ (4) \\[6pt]
\multicolumn{6}{@{}l@{}}{$s=6$, $\bolds{\beta}^{\star}=(3,-3,3,-3,3,-3)^T$} \\
\quad 0 & 100 & 15,\ (43) & 18,\ (47) & 6,\ (0) & 6,\ (1) \\
\quad 0.2 & 100 & 42,\ (116) & 47,\ (99) & 7,\ (1) & 7,\ (1) \\
\quad 0.4 & 100 & 143,\ (207) & 129,\ (226) & 12,\ (4) & 12,\ (5) \\
\quad 0.6 & 200 & 47,\ (93) & 49,\ (110) & 7,\ (1) & 7,\ (1) \\
\quad 0.8 & 200 & 360,\ (470) & 376,.5\ (486) & 54,\ (32) & 51,\ (25) \\[6pt]
\multicolumn{6}{@{}l@{}}{$s=12$, $\bolds{\beta}^{\star}=(0.6,0.8,\ldots)^T$} \\
\quad 0 & 200 & 151,\ (212) & 140,\ (207) & 15,\ (4) & 15,\ (4) \\
\quad 0.2 & 100 & 37,.5\ (10) & 36,\ (12) & 16,\ (3) & 16,\ (4) \\
\quad 0.4 & 100 & 39,\ (7) & 40,.5\ (8) & 18,\ (3) & 17,\ (2) \\
\quad 0.6 & 100 & 41,\ (7) & 42,\ (6) & 19,\ (3) & 18,\ (3) \\
\quad 0.8 & 100 & 44,\ (5) & 46,\ (6) & 23,\ (1478) & 24,\ (50) \\[6pt]
\multicolumn{6}{@{}l@{}}{$s=24$, $\bolds{\beta}^{\star}=(3,4,\ldots)^T$}\\
\quad 0 & 400 & 229,\ (283) & 227,\ (279) & 24,\ (0) & 25,\ (0) \\
\quad 0.2 & 100 & 61,\ (43) & 67,\ (46) & 30,\ (2) & 30,\ (2) \\
\quad 0.4 & 100 & 48,\ (2) & 47,\ (2) & 31,\ (2) & 30,\ (1) \\
\quad 0.6 & 100 & 48,\ (2) & 49,\ (2) & 32,\ (2) & 32,\ (3) \\
\quad 0.8 & 100 & 49,\ (2) & 49,\ (1) & 32,\ (2) & 32,\ (2) \\
\hline
\end{tabular*}     \vspace*{6pt}
\end{table}

%s8 ###
\section{Concluding remarks}\label{sec8}
In this paper, we propose two independent screening methods by
ranking the maximum marginal likelihood estimators and the maximum
marginal likelihood in generalized linear models. With
\citet{FanLv08} as a special case, the proposed method is shown to
possess the sure independence screening property. The success of the marginal
screening generates the idea that any surrogates screening, besides
the marginal utility screening introduced in this paper, as long
as which can preserve the nonsparsity structure of the true model
and is feasible in computation, can be a good option for population
variable screening. It also paves the way for the sample variable
screening, as long as the surrogate signals are uniformly
distinguishable from the stochastic noise. Along this line, many
statistics, such as R square statistics, marginal pseudo
likelihood (least square estimation, for example), can be a
potential basis for the independence learning. Meanwhile the
proposed properties of sure screening and vanishing false selection
rate will be good criteria for evaluating ultrahigh-dimensional
variable selection methods.

As our current results only hold when the log-likelihood function
is concave in the regression parameters, the proposed procedure
does not cover all generalized linear models, such as some
noncanonical link cases. This leaves space for future research.

Unlike \citet{FanLv08}, the main idea of our technical proofs is
broadly applicable. We conjecture that our results should hold
when the conditional distribution of the outcome $Y$ given the
covariates $\mathbf{X}$ depends only on $\mathbf{X}^T \bolds{\beta
}^{\star}$ and is
arbitrary and
unknown otherwise. Therefore, besides GLIM, the SIS method can be
applied to a rich class of general regression models, including
transformation models [\citet{BoxCox64}; \citet{Bickel81}], censored
regression models [\citet{Cox72}; \citet{Koso04};
\citet{Zeng07}] and projection pursuit
regression [\citet{Friedman81}]. These are also interesting future
research topics.

Another important extension is to generalize the concept of
marginal regression to the marginal group regression, where the
number of covariates $m$ in each marginal regression is greater or
equal to one. This leads to a new procedure called grouped\vadjust{\goodbreak}
variables screening. It is expected to improve the situation when the
variables are highly correlated and jointly important, but marginally
the correlation between each individual variable and the response is
weak. The current theoretical studies for the
componentwise marginal regression can be directly extended to
group variable screening, with appropriate conditions and
adjustments. This leads to another interesting topic of future research.

In practice, how to choose the tuning parameter $\gamma_n$ is an interesting
and important problem. As discussed in \citet{FanLv08},
for the first stage of the iterative SIS procedure, our preference is
to select sufficiently many features, such that $|\mathcal{M}_{\gamma
_n}|=n$ or $n/\log(n)$. The FDR-based methods in multiple comparison
can also possibly employed. In the second
or final stage, Bayes information type of criterion can be applied. In
practice, some data-driven methods may also be welcome for choosing the
tuning parameter $\gamma_n$.
This is an interesting future research topic and is beyond the scope of
the current paper.

%s9 ###
\section{Proofs}\label{sec9}
To establish Theorem~\ref{the-1}, the following symmetrization
theorem in \citet{vw96}, contraction theorem in
Ledoux and Talagrand (\citeyear{Ledo91}) and concentration theorem in Massart (\citeyear{Massart00}) will
be needed. We reproduce them here for the sake of readability.
%l2
\begin{lemma}[{[Symmetrization, Lemma 2.3.1, van der Vaart
and Wellner (\citeyear{vw96})]}]\label{lem-2}
Let $Z_1, \ldots, Z_n$ be independent random variables with values
in $\mathcal{Z}$ and $\mathcal{F}$ is a class of real valued
functions on $\mathcal{Z}$. Then
\[
E \Bigl\{ \sup_{f \in\mathcal{F}} |(\mathbb{P}_n - P)f(Z) | \Bigr\}\le
2E \Bigl\{\sup_{f \in\mathcal{F}}|\mathbb{P}_n \varepsilon f(Z)| \Bigr\},
\]
where $\varepsilon_1, \ldots, \varepsilon_n$ be a Rademacher
sequence (i.e., i.i.d. sequence taking values $\pm1$ with
probability $1/2$) independent of $Z_1, \ldots, Z_n$ and $Pf(Z) = E
f(Z)$.
\end{lemma}
%l3
\begin{lemma}[{[Contraction theorem \citet{Ledo91}]}]\label{lem-3}
Let $z_1,\break \ldots, z_n$ be nonrandom elements of some space
$\mathcal{Z}$, and let $\mathcal{F}$ be a class of real valued
functions on $\mathcal{Z}$. Let $\varepsilon_1, \ldots,
\varepsilon_n$ be a Rademacher sequence. Consider Lipschitz
functions $\gamma_i\dvtx  \mathbf{R} \mapsto\mathbf{R}$, that is,
\[
|\gamma_i(s) - \gamma_i(\tilde s)| \le|s - \tilde s|\qquad
\forall s, \tilde s \in\mathbf{R}.
\]
Then for any function $\tilde f\dvtx  \mathcal{Z} \mapsto\mathbf{R}$,
we have
\[
E\Bigl\{\sup_{f \in\mathcal{F}}\big| \mathbb{P}_n \varepsilon\bigl(\gamma(f) -
\gamma(\tilde f)\bigr)\big| \Bigr\}
\le2 E\Bigl\{\sup_{f \in\mathcal{F}}| \mathbb{P}_n \varepsilon(f -
\tilde f)| \Bigr\}.
\]
\end{lemma}
%l4
\begin{lemma}[{[Concentration theorem \citet{Massart00}]}]\label{lem-4}
Let $Z_1,\ldots, Z_n$ be independent random variables with values
in some space $\mathcal{Z}$ and let $\gamma\in\Gamma$, a class
of real valued functions on $\mathcal{Z}$. We assume that for some
positive constants $l_{i,\gamma}$ and $u_{i,\gamma}$,
$ l_{i,\gamma} \le\gamma(Z_i) \le u_{i,\gamma}$ $\forall\gamma\in
\Gamma$.
Define
\[
L^2 = \sup_{\gamma\in\Gamma} \sum_{i=1}^n
(u_{i,\gamma}-l_{i,\gamma})^2/n
\]
and
\[
\mathbf{Z} = \sup_{\gamma\in\Gamma} |(\mathbb{P}_n - P) \gamma
(Z) |,
\]
then for any $t >0$,
\[
P (\mathbf{Z} \ge E\mathbf{Z} + t ) \le
\exp\biggl(-\frac{n t^2}{2L^2} \biggr).
\]
\end{lemma}

%t9 ###
%
\begin{table}
\caption{The MMMS and the associated RSD (in the parenthesis) of the
simulated examples~for~linear regressions in Setting 3 (\textup{S3}), where
$p=2000$ and $n=600$. The~values with italic font have the same meaning as
Table~\protect\ref{table-2}. M-$\lambda_{\max}$ and its~RSD~have~the~same~meaning as
Table~\protect\ref{table-1}}\label{table-9}
\begin{tabular*}{\textwidth}{@{\extracolsep{4in minus 4in}}l e{2.8} e{2.4}e{2.4} e{4.4} e{4.3}@{}}
\hline
$\bolds{s}$ & \multicolumn{1}{c}{$\bolds{M}\textbf{-}\bolds{\lambda_{\max}}$\textbf{(RSD)}}
& \multicolumn{1}{c}{\textbf{SIS-MLR}} & \multicolumn{1}{c}{\textbf{SIS-MMLE}}& \multicolumn{1}{c}{\textbf{LASSO}}&
\multicolumn{1}{c@{}}{\textbf{SCAD}}\\
\hline
\phantom{0}3 & 8,.47\ (0.17)& 3,\ (0) & 3,\ (0) & 3,\ (0) & 3,\ (0) \\
\phantom{0}6 & 10,.36\ (0.26)& 56,\ (0) & 56,\ (0) & 47,\ (4) & 45,\ (3) \\
12           & 14,.69\ (0.39)& 62,\ (0) & 62,\ (0) & \mathit{1610},\ \mathit{(10)} & \mathit{1304},\ \mathit{(2)}
\\
24&23,.70\ (0.14)&81,\ (19) & 81,\ (23) & \mathit{1637},\ \mathit{(14)} & \mathit{1303},\ \mathit{(1)}\\
\hline
\end{tabular*}
\end{table}

Let $N > 0$, define a set of $\bolds{\beta}$
\[
\mathcal{B}(N) = \{\bolds{\beta}\in\mathcal{B},\|\bolds{\beta}-
\bolds{\beta}_0\| \le N \}.
\]
Let
\[
\mathbb{G}_1(N) = \sup_{\bolds{\beta}\in\mathcal{B}(N)}
|(\mathbb{P}_n - P)\{l(\mathbf{X}^T \bolds{\beta}, Y) - l(\mathbf
{X}^T \bolds{\beta}_0,
Y)\}I_n(\mathbf{X}, Y) |,
\]
where $I_n(\mathbf{X}, Y)$ is defined in condition~B.
The next result is about the upper bound of the tail probability
for $\mathbb{G}_1(N)$ in the neighborhood of $\mathcal{B}(N)$.
%l5
\begin{lemma}\label{lem-5}
For all $t>0$, it holds that
\[
P\bigl(\mathbb{G}_1(N) \ge4Nk_n (q /n)^{1/2} (1+t) \bigr) \le
\exp( - 2 t^2 / K_n^2).
\]
\end{lemma}

\begin{pf}
The main idea is to apply the
concentration theorem (Lemma~\ref{lem-4}). To this end, we first
show that the random variables involved are bounded. By condition~B and the Cauchy--Schwarz inequality, we have that on the
set~$\Omega_n$,
\[
| l(\mathbf{X}^T \bolds{\beta}, Y) - l(\mathbf{X}^T \bolds{\beta}_0,
Y)| \leq k_n | \mathbf{X}^T (\bolds{\beta}- \bolds{\beta}_0)|
\leq k_n \|\mathbf{X}\| \| \bolds{\beta}- \bolds{\beta}_0\|.
\]
On the set $\Omega_{n}$, by the definition of $\mathcal{B}(N)$,
the above random variable is further bounded by $k_n q^{1/2} K_n N$.
Hence, $L^2 = 4 k_n^2 q K_n^2 N^2$, using the notation of
Lemma~\ref{lem-4}.

We need to bound the expectation $E \mathbb{G}_1(N)$.
An application of the symmetrization theorem (Lemma~\ref{lem-2})
yields that
\begin{eqnarray*}
E \mathbb{G}_1(N) \le2 E \Bigl[\sup_{\bolds{\beta}\in\mathcal{B}(N)}
|\mathbb{P}_n \varepsilon\{l(\mathbf{X}^T \bolds{\beta}, Y) -
l(\mathbf{X}^T
\bolds{\beta}_0, Y) \} I_n(\mathbf{X}, Y) | \Bigr].
\end{eqnarray*}
By the contraction theorem (Lemma~\ref{lem-3}), and the Lipschitz
condition in condition~B, we can bound the right-hand side of the
above inequality further by
%e6 ###
%
\begin{eqnarray}\label{e5}
4 k_n E \Bigl\{\sup_{\bolds{\beta}\in\mathcal{B}(N)} |\mathbb{P}_n
\varepsilon
\mathbf{X}^T ( \bolds{\beta}- \bolds{\beta}_0) I_n(\mathbf{X}, Y)
| \Bigr\}.
\end{eqnarray}
By the Cauchy--Schwarz inequality, the expectation in (\ref{e5})
is controlled by
%e7 ###
%
\begin{eqnarray}\label{e6}
E \| \mathbb{P}_n \varepsilon\mathbf{X}I_n(\mathbf{X}, Y)\|
\sup_{\bolds{\beta}\in\mathcal{B}(N)} \|\bolds{\beta}- \bolds
{\beta}_0 \|
\leq E \| \mathbb{P}_n \varepsilon\mathbf{X}I_n(\mathbf{X}, Y) \| N.
\end{eqnarray}
By Jensen's inequality, the expectation in (\ref{e6}) is bounded
above by
\begin{eqnarray*}
(E \| \mathbb{P}_n \varepsilon\mathbf{X}I_n(\mathbf{X}, Y) \|^2
)^{1/2} =
\bigl( E \| \mathbf{X}\|^2 I_n(\mathbf{X}, Y) / n \bigr)^{1/2}
\leq(q /n)^{1/2},
\end{eqnarray*}
by noticing that
\[
E \| \mathbf{X}\|^2 I_n(\mathbf{X}, Y) \le
E \| \mathbf{X}\|^2 = E(X_1^2 + \cdots+ X_q^2) = q,
\]
since $EX_j^2 = 1$. Combining these results, we conclude that
\[
E \mathbb{G}_1(N) \leq
4 N k_n (q / n)^{1/2}.
\]

An application of the concentration theorem (Lemma~\ref{lem-4})
yields that
\begin{eqnarray*}
P\bigl(\mathbb{G}_1(N) \ge4Nk_n (q /n)^{1/2} (1+t) \bigr) & \le&
\exp\biggl(- \frac{n\{4Nk_n (q /n)^{1/2} t\}^2}{8q K_n^2 k_n ^2
N^2 } \biggr) \\
& = & \exp( - 2 t^2 / K_n^2).
\end{eqnarray*}
This proves the lemma.\vadjust{\goodbreak}
\end{pf}

\begin{pf*}{Proof of Theorem~\ref{the-1}}
The proof takes two main steps:
we first bound $\|\hat{\bolds{\beta}} - \bolds{\beta}_0 \|$ by
$\mathbb{G}(N)$ for
a small $N$, where $N $ chosen so that conditions~B and~C hold, and
then utilize Lemma~\ref{lem-5} to
conclude.

Following a similar idea in \citet{Geer02}, we define a convex
combination $\bolds{\beta}_{s} = s \hat{\bolds{\beta}} + (1-s)
\bolds{\beta}_0$ with
\[
s = (1 + \| \hat{\bolds{\beta}} - \bolds{\beta}_0\|/N )^{-1}.
\]
Then, by definition,
\[
\|\bolds{\beta}_{s} - \bolds{\beta}_0\| =
s\| \hat{\bolds{\beta}} - \bolds{\beta}_0\| \le N,
\]
namely, $\bolds{\beta}_{s} \in\mathcal{B}(N)$. Due to the
convexity, we
have
%e8 ###
%
\begin{eqnarray}\label{e7}
\mathbb{P}_n l(\mathbf{X}^T \bolds{\beta}_{s}, Y) &\le& s \mathbb
{P}_n l(\mathbf{X}^T
\hat{\bolds{\beta}}, Y)
+ (1-s) \mathbb{P}_n l(\mathbf{X}^T \bolds{\beta}_0, Y) \nonumber\\[-8pt]\\[-8pt]
&\le& \mathbb{P}_n l(\mathbf{X}^T \bolds{\beta}_0, Y).\nonumber
\end{eqnarray}

Since $\bolds{\beta}_0$ is the minimizer, we have
\[
E[l(\mathbf{X}^T
\bolds{\beta}_s, Y) - l(\mathbf{X}^T \bolds{\beta}_0, Y)] \geq0,
\]
where $\bolds{\beta}_s$
is regarded a parameter in the above expectation. Hence, it
follows from (\ref{e7}) that
\begin{eqnarray*}
& & E[l(\mathbf{X}^T \bolds{\beta}_s, Y) -
l(\mathbf{X}^T \bolds{\beta}_0, Y)]
\nonumber\\
&&\qquad \le (E-\mathbb{P}_n) [l(\mathbf{X}^T \bolds{\beta}_s, Y) -
l(\mathbf{X}^T \bolds{\beta}_0, Y)] \nonumber\\
&&\qquad \leq\mathbb{G}(N),
\end{eqnarray*}
where
\[
\mathbb{G}(N) = \sup_{\bolds{\beta}\in\mathcal{B}(N)}
|(\mathbb{P}_n - P)\{l(\mathbf{X}^T \bolds{\beta}, Y) - l(\mathbf
{X}^T \bolds{\beta}_0,
Y)\}|.
\]

By condition~C, it follows that
%e9 ###
%
\begin{eqnarray} \label{e8}
\| \bolds{\beta}_s - \bolds{\beta}_0 \| \leq[\mathbb{G}(N)/V_n]^{1/2}.
\end{eqnarray}

We now use (\ref{e8}) to conclude the result. Note that for any
$x$,
\[
P( \| \bolds{\beta}_s - \bolds{\beta}_0 \| \geq x) \leq P\bigl( \mathbb
{G}(N) \geq V_n
x^2\bigr).
\]
Setting $x = N/2$, we have
\[
P ( \| \bolds{\beta}_s - \bolds{\beta}_0 \| \geq N/2) \leq
P \{ \mathbb{G}(N) \geq V_n N^2/4 \}.
\]
Using the definition of $\bolds{\beta}_s$, the left-hand side is the
same as $P \{ \| \hat{\bolds{\beta}} - \bolds{\beta}_0 \| \geq N \}
$. Now, by
taking $N = 4 a_{n}(1+t)/V_n$ with $a_n = 4 k_n \sqrt{q /n}$,
we have
\begin{eqnarray*}
P \{ \| \hat{\bolds{\beta}} - \bolds{\beta}_0 \| \geq N \} & \leq&
P\{ \mathbb{G}(N) \geq V_n N^2/4 \} \\
&=& P \{ \mathbb{G}(N) \geq N a_n (1+t) \}.
\end{eqnarray*}
The last probability is bounded by
%e10 ###
%
\begin{equation}
P \{ \mathbb{G}(N) \geq N a_n (1+t), \Omega_{n, \star} \} +
P \{ \Omega_{n, \star}^c \}, \label{e9}
\end{equation}
where $\Omega_{n, \star} = \{\|\mathbf{X}_i\| \leq K_n, |Y_i| \leq
K_n^\star\}$.

On the set $\Omega_{n, \star}$, since
\[
\sup_{\bolds{\beta}\in\mathcal{B}(N)}\mathbb{P}_n | l(\mathbf
{X}^T \bolds{\beta}, Y)
- l(\mathbf{X}^T \bolds{\beta}_0, Y) | \bigl(1 - I_n(\mathbf{X}, Y)\bigr)=0,
\]
by the triangular inequality,
\[
\mathbb{G}(N) \le\mathbb{G}_1(N) + \sup_{\bolds{\beta}\in
\mathcal{B}(N)}
\big| E [l(\mathbf{X}^T \bolds{\beta}, Y) - l(\mathbf{X}^T \bolds
{\beta}_0, Y)] \bigl(1 - I_n(\mathbf{X}, Y)\bigr) \big|.
\]
It follows from condition~B that (\ref{e9}) is bounded by
\[
P \{ \mathbb{G}_1(N) \geq N a_n (1+t) + o(q/n) \} +
n P \{ (\mathbf{X}, Y) \in\Omega_{n}^c \}.
\]
The conclusion follows from Lemma~\ref{lem-5}.
\end{pf*}

%%%---------------------------------------------------------------------------------------

\begin{pf*}{Proof of Theorem~\ref{the-2}}
First of all, the target
function $E l (\beta_0 + \beta_j X_j, Y)$ is a convex function in
$\beta_j$. We first show that if $\operatorname{cov}(b'(\mathbf{X}^T
\bolds{\beta}^{\star}),
X_j)=0$, then $\beta_{j}^M$ must be zero. Recall $EX_j=0$. The
score equation of the marginal regression at $\beta_j^M$ takes the
form
\begin{eqnarray*}
E \{ b'(\mathbf{X}_j^T \bolds{\beta}_{j}^M) X_j \}= E(YX_j)= E \{
b'(\mathbf{X}^T \bolds{\beta}^{\star}) X_j \} .
\end{eqnarray*}
It can be equivalently written as
\begin{eqnarray*}
\operatorname{cov}(b'(\mathbf{X}_j^T \bolds{\beta}_j^M), X_j) =
\operatorname{cov}(b'(\mathbf{X}^T \bolds{\beta}^{\star}), X_j)=0.
\end{eqnarray*}
Since both functions $f(t) = b'(\beta_{j,0}^M + t)$ and $h(t)=t$ are strictly
monotone in $t$, when $t \neq0$,
\[
\{f(t)-f(0) \}(t-0)>0.
\]

If $\beta_{j}^M \neq0,$ let $t=\beta_j^MX_j$,
\begin{eqnarray*}
\beta_{j}^M \operatorname{cov}(f(\beta_j^MX_j), X_j) = E[E \{
f(t)-f(0)\}(t-0)
|X_j\neq0]>0,
\end{eqnarray*}
which leads to a contradiction. Hence $\beta_{j}^M $ must be zero.

On the other side, if $\beta_{j}^M = 0$, the score equations now
take the form
%e12 ###
%e11 ###
%
\begin{eqnarray}\label{e10}
E \{ b'( \beta_{j,0}^M) \}=
E \{ b'(\mathbf{X}^T \bolds{\beta}^{\star}) \}
\end{eqnarray}
and
\begin{eqnarray}\label{e11}
E \{ b'(\beta_{j,0}^M)X_j \}= E \{ b'(\mathbf{X}^T
\bolds{\beta}^{\star}) X_j \} .
\end{eqnarray}
Since $b'(\beta_{j,0}^M)$ is a constant, we can get the desired
result by plugging (\ref{e10}) into~(\ref{e11}).
\end{pf*}

\begin{pf*}{Proof of Theorem~\ref{the-3}}
We first prove the case that
$b''(\theta)$ is bounded. By the Lipschitz continuity of the
function $b'(\cdot)$, we have
\begin{eqnarray*}
|\{b'(\beta_{j,0}^M + X_j \beta_{j}^M) -
b'(\beta_{j,0}^M)\}X_j | \le D_1 |\beta_{j}^M|X_j^2.
\end{eqnarray*}
$D_1 = \sup_{x} b''(x)$. By taking the expectation on both sides,
we have
\begin{eqnarray*}
|E \{b'(\beta_{j,0}^M + X_j \beta_{j}^M) - b'(\beta_{j,0}^M) \} X_j
| \le D_1 |\beta_{j}^M|,
\end{eqnarray*}
namely,
%e13 ###
%
\begin{equation}
D_1 |\beta_{j}^M| \geq| \operatorname{cov}(b'(\beta_{j,0}^M + X_j
\beta_{j}^M),
X_j)|. \label{e12}
\end{equation}
Note that $\beta_{j,0}^M$ and $\beta_{j}^M$ satisfy the score
equation
%e14 ###
%
\begin{equation}
E\{b'(\beta_{0,1}^M + \beta_j^M X_j) - b'(\mathbf{X}^T \bolds{\beta
}^\star)\}
X_j = 0.
\label{e13}
\end{equation}
It follows from (\ref{e12}) and $ E X_j = 0$ that
\[
|\beta_{j}^M| \geq D_1^{-1} c_1 n^{-\kappa}.
\]
The conclusion follows.

We now prove the second case. The result holds trivially if
$|\beta_{j}^M| \geq c n^{-\kappa}$ for a sufficiently large
universal constant $c$. Now suppose that $|\beta_{j}^M| \leq c_9
n^{-\kappa}$, for some positive constant $c_9$.
We will show later that $|\beta_{j,0}^M - \beta_0^M |
\leq c_{10}$ for some $c_{10} > 0$, where $\beta_0^M$ is such that
$b'(\beta_0^M) = EY$. In this case, if $|X_j| \leq
n^{\kappa}$, then the points $\beta_{j,0}^M$ and
$(\beta_{j,0}^M + X_j \beta_{j}^M)$ falls in the interval
$\beta_0^M \pm h$, independent of $j$, where $h = c_9+c_{10}$.

By the Lipschitz continuity of the function $b'(\cdot)$ in the
neighborhood around $\beta_0^M$, we have for $|X_j| \leq
n^{\kappa}$,
\begin{eqnarray*}
|\{b'(\beta_{j,0}^M + X_j \beta_{j}^M) - b'(\beta_{j,0}^M)\}X_j|
\le D_2 |\beta_{j}^M|X_j^2,
\end{eqnarray*}
where $D_2 = \max_{x \in[\beta_0^M - h, \beta_0^M + h]} b''(x)$.
By taking the expectation on both sides, it follows that
%e15 ###
%
\begin{eqnarray}\label{e14}
&& |E \{b'(\beta_{j,0}^M + X_j \beta_{j}^M) - b'(\beta_{j,0}^M) \} X_j
I(|X_j| \leq n^{\kappa})| \le D_2 |\beta_{j}^M| .
\end{eqnarray}
By using (\ref{e13}) and $EX_j = 0$, we deduce from (\ref{e14})
that
%e16 ###
%
\begin{equation}
D_2 |\beta_{j}^M| \geq|\operatorname{cov}(\mathbf{X}^T\bolds{\beta
}^\star, X_j)| - A_0 - A_1,
\label{e15}
\end{equation}
where $A_m = E |b'(\beta_{j,0}^M + X_j^m \beta_{j}^M) X_j|
I(|X_j| \geq n^{\kappa})$ for $m = 0$ and $1$. Since
$|\beta_{j,0}^M + X_j \beta_{j}^M| \leq a |X_j|$ for $|X_j| \geq
n^{\kappa}$ for a sufficiently large $n$, independent of $j$, by
the condition given in the theorem, we have
\[
A_m \leq E G(a|X_j|)^m |X_j| I(|X_j| \geq n^{\kappa}) \leq d
n^{-\kappa}\qquad \mbox{for $m=0$ and $1$}.
\]
The conclusion now follows from (\ref{e15}).

It remains to show that when $|\beta_{j}^M| \leq c_9 n^{-\kappa} $
we have $|\beta_{j,0}^M -\beta_0^M | \leq c_{10}$. To this end,
let
\[
\ell(\beta_0) = E \{b(\beta_0 + \beta_j^M X_j) -
Y (\beta_0 + \beta_j^M X_j)\}.
\]
Then, it is easy to see that
\[
\ell'(\beta_0) = E b'(\beta_0 + \beta_j^M X_j) - b'(\beta_0^M).
\]
Observe that
%e17 ###
%
\begin{eqnarray} \label{e16}
| E b'(\beta_0 + \beta_j^M X_j) - b'(\beta_0)| \leq R_1 + R_2,
\end{eqnarray}
where $R_1 = \sup_{|x| \leq c_9 n^{\eta-\kappa}} |b'(\beta_0+x) -
b'(\beta_0)|$ and $ R_2 = 2 E G(a|X_j|) I(|X_j| > n^{\eta})$. Now,
$R_1 = o(1)$ due to the continuity of $b'(\cdot)$ and $R_2 = o(1)$
by the condition of the theorem. Consequently, by (\ref{e16}), we
conclude that
\[
\ell'(\beta_0) = b'(\beta_0) - b'(\beta_0^M) + o(1).
\]
Since $b'(\cdot)$ is a strictly increasing function, it is now
obvious that
\[
\ell'(\beta_0^M - c_{10}) < 0,\qquad
\ell'(\beta_0^M + c_{10}) >
0
\]
for any given $c_{10} > 0$. Hence, $|\beta_{j,0}^M - \beta_0^M| <
c_{10}$.
\end{pf*}

\begin{pf*}{Proof of Proposition~\ref{prop1}}
Without loss of
generality, assume that $g(\cdot)$ is strictly increasing and
$\rho
> 0$. Since $X$ and $Z$ are jointly normally distributed, $Z$ can be
expressed as
\[
Z = \rho X + \varepsilon,
\]
where $\rho= E(XZ)$ is the regression coefficient, and $X$ and
$\varepsilon$ are independent. Thus,
%e18 ###
%
\begin{eqnarray}
E f(Z) X = E g(\rho X) X = E [g(\rho X) - g(0)]X, \label{e17}
\end{eqnarray}
where $g(x) = E f(x + \varepsilon)$ is a strictly increasing
function. The right-hand side of (\ref{e17}) is always
nonnegative and is zero if and only if $\rho=0$.

To prove the second part, we first note that the random variable
on the right-hand side of (\ref{e17}) is nonnegative. Thus, by
the mean-value theorem, we have that
\begin{eqnarray*}
|E f(Z) X| & \geq& \inf_{|x| \leq c \rho} |g'(x)| \rho E X^2 I(|X|
\leq c).
\end{eqnarray*}
Hence, the result follows.
\end{pf*}

\begin{pf*}{Proof of Lemma~\ref{lem-1}}
By Chebyshev's inequality,
\[
P(Y \ge u) \le\exp(-s_0u)E\exp(s_0Y).
\]
Let $\theta= \mathbf{X}^T \bolds{\beta}^{\star}$.
Since $Y$ belongs to an exponential family, we have
\begin{eqnarray*}
E\{\exp(s_0Y)|\theta\} & = & \exp\bigl(b(\theta+s_0) - b(\theta)\bigr).
\end{eqnarray*}
Hence
\[
P(Y \ge u) \le\exp(-s_0u)E \exp\bigl(b(\mathbf{X}^T \bolds{\beta
}^{\star}+s_0) -
b(\mathbf{X}^T \bolds{\beta}^{\star})\bigr).
\]
Similarly we can get
\[
P(Y \le-u) \le\exp(-s_0u)E \exp\bigl(b(\mathbf{X}^T \bolds{\beta
}^{\star}-s_0) -
b(\mathbf{X}^T \bolds{\beta}^{\star})\bigr).
\]
The desired result thus follows from condition~D by letting $u = m_0
t^{\alpha}/s_0$.
\end{pf*}

%Let $K_n^{\star} = K_n^{\alpha}m_0/s_0$.
\begin{pf*}{Proof of Theorem~\ref{the-4}}
Note that condition~B is
satisfied with $k_n$ defined in Section~\ref{sec5.2}. The tail part of
condition~B can also be easily checked. In fact,
\begin{eqnarray*}
& & E [l(\mathbf{X}_j^T\bolds{\beta}_j, Y) - l(\bolds{\beta}_j^M,Y) ] \bigl(1 - I_n(\mathbf{X}_j,Y)\bigr)\\
&&\qquad \leq \big|E b(\mathbf{X}_j^T\bolds{\beta}_j)I(|X_j| \geq K_n)\big|
+ \big|E b(\mathbf{X}_j^T\bolds{\beta}_j^M)I(|X_j| \geq K_n)\big|
\\
&&\qquad\quad{}+ B(\bolds{\beta}_j) + B(\bolds{\beta}_j^M),
\end{eqnarray*}
where $B(\bolds{\beta}_j) = |E Y \mathbf{X}_j^T \bolds{\beta}_j (1
- I_n(\mathbf{X}_j, Y))|$. The
first two terms are of order $o(1/n)$ by assumption, and the last two
terms can be bounded by the exponential tail conditions in condition~D and the Cauchy--Schwarz inequality.

By Theorem~\ref{the-1}, we have for any $t
> 0$,
\[
P\bigl(\sqrt{n} |\hat{\beta}_j^M - \beta_j^M| \geq16 (1 + t) k_n / V\bigr)
\leq
\exp( - 2t^2/K_n^2) + nm_1 \exp(-m_0 K_n^\alpha).
\]
By taking $1+t = c_3 V n^{1/2-\kappa}/(16k_n)$, it follows that
\[
P(|\hat{\beta}_j^M - \beta_j^M| \geq c_3 n^{-\kappa}) \leq\exp\bigl(- c_4 n^{1-2\kappa} /(k_nK_n)^2\bigr) + nm_1 \exp(-m_0 K_n^\alpha).
\]
The first result follows from the union bound of probability.

To prove the second part, note that on the event
\[
A_n \equiv\Bigl\{\max_{j \in\mathcal{M}_\star} |\hat{\beta}_j^M -
\beta_j^M| \leq c_2 n^{-\kappa}/2\Bigr\},
\]
by Theorem~\ref{the-3}, we have
%e19 ###
%
\begin{equation} \label{e18}
|\hat{\beta}_j^M| \geq c_2 n^{-\kappa}/2
\qquad\mbox{for all $j \in\mathcal{M}_\star$}.
\end{equation}
Hence, by the choice of $\gamma_n$, we have $\mathcal{M}_\star
\subset\widehat{\mathcal{M}}_{\gamma_n}$. The result now follows
from a simple union bound
\[
P(A_n^c) \leq s_n \bigl\{\exp\bigl( - c_4 n^{1-2\kappa} /(k_nK_n)^2\bigr) + nm_1
\exp(-m_0 K_n^\alpha)\bigr\}.
\]

This completes the proof.
\end{pf*}

\begin{pf*}{Proof of Theorem~\ref{the-5}}
The key idea of the proof is to show that
%e20 ###
%
\begin{equation} \label{e19}
\| \bolds{\beta}^M \|^2 = O(\| \bolds{\Sigma}\bolds{\beta}^\star
\|^2) =
O\{\lambda_{\max} (\bolds{\Sigma})\}.
\end{equation}
If so, the number of $\{j\dvtx  |\beta_j^M| > \varepsilon
n^{-\kappa}\}$ cannot exceed
$O\{n^{2\kappa}\lambda_{\max}
(\bolds{\Sigma})\}$
for any $\varepsilon> 0$. Thus, on the set
\[
B_n = \Bigl\{\max_{1 \leq j \leq p_n} |\hat{\beta}_j^M - \beta_j^M| \leq
\varepsilon
n^{-\kappa}\Bigr\},
\]
the number of $\{j\dvtx  |\hat{\beta}_j^M| > 2 \varepsilon n^{-\kappa}\}$ cannot exceed the number of $\{j\dvtx  |\beta_j^M|>\varepsilon n^{-\kappa}
\}$, which is bounded by $O\{n^{2\kappa}\lambda_{\max} (\bolds
{\Sigma})\}$.
By taking $\varepsilon=c_5/2$, we have
\[
P [ | \widehat{\mathcal{M}}_{\gamma_n}| \leq
O\{n^{2\kappa}\lambda_{\max} (\bolds{\Sigma})\} ] \geq P(B_n).
\]
The conclusion follows from Theorem~\ref{the-4}(i).

It remains to prove (\ref{e19}). We first bound $\beta_{j}^M$.
Since $b'(\cdot)$ is monotonically increasing, the function
\begin{eqnarray*}
\{b'(\beta_{j,0}^M + X_j \beta_{j}^M) - b'(\beta_{j,0}^M)\} X_j
\beta_{j}^M
\end{eqnarray*}
is always positive. By Taylor's expansion, we have
\[
\{b'(\beta_{j,0}^M + X_j \beta_{j}^M) - b'(\beta_{j,0}^M)\}
\beta_{j}^M X_j \geq D_3 (\beta_{j}^M X_j)^2 I(|X_j| \leq K),
\]
where $D_3 = \inf_{|x| \leq K(B+1)} b''(x)$, since
$(\beta_{j,0}^M, \beta_j^M)$ is an interior point of the square
$\mathcal B$ with length $2B$. By taking the expectation on both sides
and using $EX_j = 0$, we have
\[
E b'(\beta_{j,0}^M + X_j \beta_{j}^M) \beta_{j}^M X_j \geq D_3
E (\beta_{j}^M X_j)^2 I(|X_j| \leq K).
\]
Since $E X_j^2 I(|X_j| \leq K) = 1 - E X_j^2 I(|X_j| > K) $, it is
uniformly bounded from below for a sufficiently large $K$, due to the
uniform exponential tail bound in condition~D. Thus, it follows from
(\ref{e11}) that
%e21 ###
%
\begin{equation} \label{e20}
|\beta_j^M|^2 \leq D_4 |E b'(\mathbf{X}^T \bolds{\beta}^\star) X_j|
\end{equation}
for some $D_4 > 0$.

We now further bound from above the right-hand side of (\ref{e20})
by using $\operatorname{var}(\mathbf{X}^T \bolds{\beta}^{\star}) =
O(1)$. We first show the
case where $b''(\cdot)$ is bounded. By the Lipschitz continuity
of the function $b'(\cdot)$, we have
\begin{eqnarray*}
|\{b'(\mathbf{X}^T \bolds{\beta}^{\star}) - b'(\beta_{0}^{\star
})\}X_j
| \le D_5 | X_j \mathbf{X}_M^T \bolds{\beta}_1^{\star} |,
\end{eqnarray*}
where $\mathbf{X}_M = (X_1,\ldots,X_{p_n})^T$ and $\bolds{\beta
}_1^{\star} =
(\beta_1^{\star}, \ldots, \beta_{p_n}^{\star})^T$.

By putting the above equation into the vector form and taking the
expectation on both sides, we have
%e22 ###
%
\begin{eqnarray} \label{e21}
\|E \{b'(\mathbf{X}^T \bolds{\beta}^{\star}) - b'(\beta_{0}^{\star
}) \}
\mathbf{X}_M \|^2 & \le&
D_5^2 \|E \mathbf{X}_M \mathbf{X}_M^T \bolds{\beta}_1^{\star} \|
^2\nonumber\\[-8pt]\\[-8pt]
& \leq& D_5^2 \lambda_{\max}(\bolds{\Sigma})
\|\bolds{\Sigma}^{1/2} \bolds{\beta}^{\star} \|^2. \nonumber
\end{eqnarray}
Using $E\mathbf{X}_M = 0$ and $\operatorname{var}(\mathbf{X}^T
\bolds{\beta}^{\star}) = O(1)$,
we conclude that
\[
\|E b'(\mathbf{X}^T \bolds{\beta}^{\star})\mathbf{X}_M \|^2 \le D_6
\lambda_{\max}(\bolds{\Sigma})
\]
for some positive constant $D_6$. This together with (\ref{e20})
entails (\ref{e19}).

It remains to bound (\ref{e20}) for the second case. Since $\mathbf
{X}_M =
R \bolds{\Sigma}_1^{1/2} \mathbf{U}$, it follows that
\[
E b'(\beta_{0}^{\star} + \mathbf{X}_M^T \bolds{\beta}_1^{\star})
\mathbf{X}_M = E
b'(\beta_{0}^{\star} + \bolds{\beta}_2^T R \mathbf{U}) R \bolds
{\Sigma}_1^{1/2} \mathbf{U},
\]
where $\bolds{\beta}_2 = \bolds{\Sigma}_1^{1/2} \bolds{\beta
}_1^{\star}$. By conditioning
on $\bolds{\beta}_2^T\mathbf{U}$, it can be computed that
\[
E (\mathbf{U}| \bolds{\beta}_2^T \mathbf{U}) = \bolds{\beta}_2^T
\mathbf{U}/ \|\bolds{\beta}_2\|^2
\bolds{\beta}_ 2.
\]
Therefore,
\begin{eqnarray*}
E b'(\beta_{0}^{\star} + \mathbf{X}_M^T \bolds{\beta}_1^{\star})
\mathbf{X}_M & = & E
b'(\beta_{0}^{\star} + \bolds{\beta}_2^T R \mathbf{U}) R \bolds
{\Sigma}_1^{1/2}
\bolds{\beta}_2^T
\mathbf{U}/ \|\bolds{\beta}_2\|^2 \bolds{\beta}_2 \\
& = & E b'(\mathbf{X}^T \bolds{\beta}^\star) (\mathbf{X}^T \bolds
{\beta}^\star-
\bolds{\beta}_0^\star) \bolds{\Sigma}_1^{1/2} \bolds{\beta}_2 /
\|\bolds{\beta}_2\|^2.
\end{eqnarray*}
This entails that
%e23 ###
%
\begin{equation}\label{e22}
\hspace*{10pt}\| E b'(\mathbf{X}^T \bolds{\beta}^{\star}) \mathbf{X}_M\|^2 =
|E b'(\mathbf{X}^T \bolds{\beta}^\star) (\mathbf{X}^T \bolds
{\beta}^\star- \bolds{\beta}_0)|^2
\| \bolds{\Sigma}_1^{1/2} \bolds{\beta}_2 \|^2 / \|\bolds{\beta
}_2\|^4.
\end{equation}
By condition~G, $|E b'(\mathbf{X}^T \bolds{\beta}^\star) (\mathbf
{X}^T \bolds{\beta}^\star-
\beta_0^{\star})| =O(1)$. We also observe the facts that
$\|\bolds{\Sigma}_1^{1/2}\bolds{\beta}_2\| \le\lambda_{\max
}^{1/2}(\bolds{\Sigma})\|
\bolds{\beta}_2\|$
and that $\|\bolds{\beta}_2\| = \|\bolds{\Sigma}^{1/2} \bolds
{\beta}^{\star}\|$ is
bounded. This proves (\ref{e19}) for the second case by using
(\ref{e20}) and completes the proof.
\end{pf*}

%--------------------------------------------------------------------

\begin{pf*}{Proof of Theorem~\ref{the-6}}
If $\operatorname
{cov}(b'(\mathbf{X}^T
\bolds{\beta}^{\star}),X_j)=0$, by Theorem~\ref{the-2}, we have
$\beta_j^M=0$, hence by the model identifiability at $\bolds{\beta}_0^M$,
$\beta_{j,0}^M = \beta_{0}^M$. Hence, $L_{j}^{\star}=0$. On the
other hand, if $L_{j}^{\star}=0$, by condition~C$'$, it follows that
$\bolds{\beta}_j^M =\bolds{\beta}_0^M$, that is, $\beta_{j,0}^M =
\beta_0^M$ and
$\beta_{j}^M = 0$. Hence by Theorem~\ref{the-2}, $\operatorname
{cov}(b'(\mathbf{X}^T
\bolds{\beta}^{\star}),\break X_j)=0$.
\end{pf*}

\begin{pf*}{Proof of Theorem~\ref{the-7}} If $|\operatorname
{cov}(b'(\mathbf{X}^T
\bolds{\beta}^{\star}),X_j)| \ge c_1 n^{-\kappa}$, for $j \in
\mathcal{M}_{\star}$, by Theorem~\ref{the-3}, we have
$\min_{j\in\mathcal{M}_{\star}}|\beta_j^M| \ge c_2 n^{-\kappa}$.
The first result thus follows from condition~C$'$.

To prove the second result, we will bound $L_{j}^{\star}$. We
first show the case where $b''(\cdot)$ is bounded. By definition,
we have
%e24 ###
%
\begin{equation}\label{e23}
0 \le L_j^{\star} \le E \{l(\beta_{j,0}^M, Y) - l(\mathbf{X}_j^T
\bolds{\beta}_{j}^M, Y) \}.
\end{equation}
By Taylor's expansion of the right-hand side of (\ref{e23}), we
have that
%e25 ###
%
\begin{equation}\label{e24}
E \{l(\beta_{j,0}^M, Y) - l(\mathbf{X}_j^T \bolds{\beta}_{j}^M, Y)
\}
\le D_5 (\beta_j^{M})^2\qquad\mbox{for some } D_5 > 0.
\end{equation}
The desired result thus follows from (\ref{e23}), (\ref{e24}) and
the proof in Theorem~\ref{the-5} that
\[
\|\mathbf{L}^{\star}\| \le O(\|\bolds{\beta}^M\|^2) =
O(\lambda_{\max}(\bolds{\Sigma})).
\]

Now we prove the second case. By the mean-value theorem,
%e26 ###
%
\begin{equation}\label{e25}
E \{ l(\beta_{j,0}^M, Y) - l(\mathbf{X}_j^T \bolds{\beta}_{j}^M, Y)
\} = E \{ Y - b'(\beta_{j,0}^M + sX_j\beta_j^M) \}
X_j \beta_j^M
\end{equation}
for some $0 < s < 1$. Since $EYX_j = Eb'(\mathbf{X}_j^T \bolds{\beta
}_j^M)X_j$,
the last term is equal to
%e27 ###
%
\begin{equation} \label{e26}
E \{ b'(\mathbf{X}_j^T \bolds{\beta}_j^M) - b'(\beta_{j,0}^M +
sX_j\beta_j^M) \}
X_j \beta_j^M.
\end{equation}
By the monotonicity of $b'(\cdot)$, when $X_j \beta_j^M \geq0$,
both factors in (\ref{e26}) is nonnegative, and hence
%e28 ###
%
\begin{equation} \label{e27}
\{ b'(\mathbf{X}_j^T \bolds{\beta}_j^M) - b'(\beta_{j,0}^M +
sX_j\beta_j^M) \} X_j \beta_j^M
\leq\{ b'(\mathbf{X}_j^T \bolds{\beta}_j^M) - b'(\beta_{j,0}^M) \}
X_j \beta_j^M.\hspace*{-35pt}
\end{equation}
When $X_j \beta_j^M < 0$, both factors in (\ref{e27}) are negative
and (\ref{e27}) continues to hold. It follows from
(\ref{e25})--(\ref{e27}) and $E X_j = 0$, the right-hand side of
(\ref{e25}) is bounded by
%e29 ###
%
\begin{equation}\label{e28}
E b'(\mathbf{X}_j^T \bolds{\beta}_j^M) X_j \beta_j^M
= E b'(\mathbf{X}^T \bolds{\beta}^{\star}) X_j \beta_j^M.
\end{equation}
Combining (\ref{e23}), (\ref{e25}) and (\ref{e28}), we can bound
$\|\mathbf{L}^{\star} \|$ in the vector form by the Cauchy--Schwarz
inequality
\[
\|\mathbf{L}^{\star} \| \le\| E b'(\mathbf{X}^T \bolds{\beta
}^{\star}) \mathbf{X}_M
\| \|\bolds{\beta}^M\| = O(\|\bolds{\Sigma}\bolds{\beta}^{\star}
\| \| \bolds{\beta}^M\|),
\]
where (\ref{e22}) is used in the last equality. The desired result
thus follows from Theorem~\ref{the-5}.
\end{pf*}

\begin{pf*}{Proof of Theorem~\ref{the-8}}
To prove the result, we first
bound $L_{j,n}$ from below to show the strength of the signals.
Let $\hat{\bolds{\beta}}{}^M_0 = (\hat{\beta}_0^M, 0)^T$. Then, by
Taylor's expansion,
we have
%e30 ###
%
\begin{equation} \label{e29}
2 L_{j,n} = (\hat{\bolds{\beta}}{}^M_0 - \hat{\bolds{\beta}}{}^M_j)\ell_j''(\bolds{\xi}_n)
(\hat{\bolds{\beta}}{}^M_0 - \hat{\bolds{\beta}}{}^M_j) \geq\lambda_{j,\min}(\hat{\beta}{}^M_j)^2,
\end{equation}
where $\lambda_{j, \min}$ is the minimum eigenvalue of the Hessian
matrix
\[
\ell_j''(\bolds{\xi}_n) = \mathbb{P} _n b''(\bolds{\xi}_n^T
\mathbf{X}_j) \mathbf{X}_j \mathbf{X}_j^T,
\]
where $\bolds{\xi}_n$ lies between $\hat{\bolds{\beta}}{}^M_0$ and
$\hat{\bolds{\beta}}{}^M_j$. We
will show
%e31 ###
%
\begin{equation} \label{e30}
P\{ \lambda_{j,\min} > c_{11}\} =1 - O \{ \exp( - c_{12} n^{1-\kappa
})\}
\end{equation}
for some $c_{11} > 0$ and $c_{12} > 0$.

Suppose (\ref{e30}) holds. Then, by (\ref{e18}), we have
\begin{eqnarray*}
& & P\Bigl\{\min_{j \in\mathcal{M}_\star} |\hat{\beta}_j^M| \geq c_2
n^{-\kappa}/2\Bigr\} \\
&&\qquad =  1 - O \bigl( s_n \bigl\{\exp\bigl( - c_4 n^{1-2\kappa} /(k_nK_n)^2\bigr) +
nm_1 \exp(-m_0 K_n^\alpha)\bigr\} \bigr).
\end{eqnarray*}
This, together with (\ref{e29}) and (\ref{e30}), implies
\begin{eqnarray*}
& & P\Bigl\{\min_{j \in\mathcal{M}_\star} L_{j,n} \geq c_{11} c_2^2
n^{-2\kappa}/8\Bigr\} \\
&&\qquad =  1 - O \bigl( s_n \bigl\{\exp\bigl( - c_4 n^{1-2\kappa} /(k_nK_n)^2\bigr) +
nm_1 \exp(-m_0 K_n^\alpha)\bigr\} \bigr).
\end{eqnarray*}
Hence, by choosing the thresholding $\nu_n = c_7 n^{-2 \kappa}$, for
$c_7 < c_{11} c_2^2/8$, $\mathcal{M}_\star\subset\widehat{\mathcal{N}}_{\nu_n}$ with the probability tending to one exponentially fast,
and the result follows.

We now prove (\ref{e30}). It is obvious that
\[
\ell_j''(\bolds{\xi}_n) \geq\min_{|x|\leq(B+1)K} b''(x) \mathbb
{P} _n \mathbf{X}_j
\mathbf{X}_j^T I(|X_j| \leq
K)
\]
for any given $K$. Since the random variable involved is
uniformly bounded in $j$, it follows from the Hoeffding inequality
[\citet{Hoff63}] that
%e32 ###
%
\begin{equation} \label{e31}
P \bigl\{ \big|(\mathbb{P} _n - P) X_j^k I(|X_j| \leq K)\big| > \varepsilon\bigr\}
\leq\exp\bigl(- 2 n \varepsilon^2/(4K^{2k})\bigr)
\end{equation}
for any $k \geq0$ and $\varepsilon> 0$. By taking $\varepsilon
= n^{-\kappa/2}$, we have
\[
P \bigl\{ \big|(\mathbb{P} _n - P) X_j^k I(|X_j| \leq K)\big| > n^{-\kappa/2} \bigr\}
\leq\exp\bigl( - 2 n^{1-\kappa}/(4K^{2k})\bigr).
\]
Consequently, with probability tending to one exponentially fast,
we have
%e33 ###
%
\begin{equation}\label{e32}
\ell_j''(\bolds{\xi}_n) \geq\min_{|x|\leq(B+1)K} b''(x) E \mathbf
{X}_j \mathbf{X}_j^T
I(|X_j| \leq
K)/2.
\end{equation}
The minimum eigenvalue of the matrix $E \mathbf{X}_j \mathbf{X}_j^T
I(|X_j| \leq
K)$ is
\[
\min_{|a| \leq1} E \bigl(a^2 + 2 a \sqrt{1 - a^2} X_j + (1-a^2)
X_j^2\bigr)I(|X_j| \leq K).
\]
It is bounded from below by
%e34 ###
%
\begin{equation}\label{e33}
\hspace*{20pt}\min_{|a| \leq1} E \{a^2 + (1-a^2)X_j^2I(|X_j| \leq K)\} - 2 \big|E X_jI(|X_j| \le K)\big| - K^{-2},
\end{equation}
where we used $P(|X_j| \geq K) \leq K^{-2}$. Since $EX_j=0$ and
$EX_j^2=1$,
\[
\big|E X_jI(|X_j| \le K)\big| = \big|E X_jI(|X_j| > K)\big| \le K^{-1} EX_j^2I(|X_j|> K) \le K^{-1}.
\]
Hence the quantity in (\ref{e33}) can be further bounded from
below by
\begin{eqnarray*}
&&EX_j^2I(|X_j| \le K) + \min_{|a| \leq1} a^2 EX_j^2I(|X_j| > K) -
2K^{-1}-K^{-2}\\
&&\qquad\ge 1 - \sup_j EX_j^2I(|X_j| > K) - 2K^{-1} - K^{-2}.
\end{eqnarray*}
The result follows from condition~G and (\ref{e32}).
\end{pf*}
\begin{pf*}{Proof of Theorem~\ref{the-9}} By (\ref{e29}), it can be
easily seen that
\[
2 L_{j, n} \leq D_1 \lambda_{\max} (\mathbb{P} _n \mathbf{X}_j
\mathbf{X}_j^T) \|
\hat{\bolds{\beta}}{}^M_0 - \hat{\bolds{\beta}}{}^M_j\|^2,
\]
where $D_1 = \sup_x b''(x)$ as defined in the proof of
Theorem~\ref{the-3}. By (\ref{e31}), with the exception on a set
with negligible probability, it follows that
\[
\lambda_{\max} (\mathbb{P} _n \mathbf{X}_j \mathbf{X}_j^T) \leq2
\lambda_{\max}(E
\mathbf{X}_j \mathbf{X}_j^T) = 2,
\]
uniformly in $j$. Therefore, with probability tending to one
exponentially fast, we have
%e35 ###
%
\begin{equation} \label{e34}
L_{j,n}\leq D_7 \{(\hat{\beta}_{0}^M - \hat{\beta}_{j,0}^M)^2 +
(\hat{\beta}_{j}^M)^2 \}
\end{equation}
for some $D_7 > 0$.

We now use (\ref{e34}) to show that if $L_{j,n} > c_7 n^{-2\kappa}$,
then $|\hat{\beta}_j^M| \geq D_8 n^{-\kappa}$, with exception on a set
with negligible probability, where $D_8 = \{c_8/(2D_7)\}^{1/2}$.
This implies that
\[
|\widehat{\mathcal{N}}_{\nu_n}| \leq|\widehat{\mathcal{M}}_{\gamma_n}|,
\]
with $\gamma_n = D_8 n^{-\kappa}$. The conclusion then follows
from Theorem~\ref{the-5}.

We now show that $L_{j,n} > c_7 n^{-2\kappa}$ implies that
$|\hat{\beta}_j^M| \geq D_8 n^{-\kappa}$,
with exception on a set with
negligible probability. Suppose that $|\hat{\beta}_j^M| < D_8
n^{-\kappa}$. From the likelihood equations, we have
%e36 ###
%
\begin{equation}\label{e35}
b'(\hat{\beta}^M_0) = \bar{Y} = \mathbb{P} _n b'(\hat{\beta}_{j,0}^M + \hat{\beta}_j^M X_j).
\end{equation}
From the proof of Theorem~\ref{the-4}, with exception on a set with
negligible probability, we have $|\hat{\beta}_0^M - \beta_{0}^M| \le
c_{13} n^{-\kappa}$ and $|\hat{\beta}_{j,0}^M - \beta_{j,0}^M| \le
c_{14} n^{-\kappa}$, for some constants $c_{13}$ and $c_{14}$. Since
$(\beta_{0}^M, 0)$ and $(\beta_{j,0}^M, \beta_j^M)$ are interior
points of the square $\mathcal B$ with length $2B$, it follows that with
exception on a set with negligible probability, $|\hat{\beta}_0^M|
\le B$
and $|\hat{\beta}_{j,0}^M| \le B$. Recall $D_1 = \sup_x b''(x)$. By
Taylor's expansion, for some $0 < s < 1$, we have
%e37 ###
%
\begin{eqnarray} \label{e36}
|\mathbb{P} _n b'(\hat{\beta}_{j,0}^M + \hat{\beta}_j^M
X_j) - b'(\hat{\beta}_{j,0}^M)| &=&
|b''(\hat{\beta}_{j,0}^M + s \hat{\beta}_j^M X_j) \hat{\beta}_j^M
\mathbb{P} _n X_j|
\nonumber\\[-8pt]\\[-8pt]
& \le& D_1 |\hat{\beta}_j^M \mathbb{P} _n X_j| = o_P(|\hat{\beta}_j^M|),\nonumber
\end{eqnarray}
where the last step follows from the facts that $EX_j = 0$ and
consequently $\mathbb{P} _n X_j = o(1)$ with an exception on a set of
negligible probability, by applying the Hoeffding inequality
on $\Omega_{n}^c$ and considering the exponential tail property of $X_j$.
Hence,
by (\ref{e35}) and (\ref{e36}), we have
\[
|b'(\hat{\beta}_0^M) - b'(\hat{\beta}_{j,0}^M)| = o_P(|\hat{\beta}_j^M|).
\]
Let $D_9 = \inf_{|x| \leq2B} b''(x)$, with exception on a set with
negligible probability, we have
\[
|b'(\hat{\beta}_0^M) - b'(\hat{\beta}_{j,0}^M)| \geq D_9 |\hat
{\beta}_0^M - \hat{\beta}
_{j,0}^M|.
\]
Therefore, we conclude that
\[
|\hat{\beta}_0^M - \hat{\beta}_{j,0}^M| = o_P(|\hat{\beta}_j^M|).
\]

By (\ref{e34}), we have $|\hat{\beta}_j^M| > D_8 n^{-\kappa}$. This is
a contraction, except on a set that has a negligible
probability. This completes the proof.
\end{pf*}

%suskaldyti doi

\section*{Acknowledgments}
The bulk of the work was conducted when Rui Song was a postdoctoral
research fellow at Princeton University.
The authors would like to thank the associate editor and two referees
for their constructive
comments that improve the presentation and the results of the paper.

\printaddresses

\end{document}